\documentclass[final,10pt,twocolumn]{IEEEtran}
\usepackage{amsmath,amssymb,graphicx,psfrag,cite,url}
\usepackage[caption=false,font=footnotesize]{subfig}

\newcommand{\bc}{\begin{center}}
\newcommand{\ec}{\end{center}}

\newcommand{\ba}{\begin{array}}
\newcommand{\ea}{\end{array}}

\newcommand{\bt}{\begin{tabular}}
\newcommand{\et}{\end{tabular}}

\newcommand{\eems}{\end{multline*}}
\newcommand{\eem}{\end{multline}}
\newcommand{\eeams}{\end{eqnarray}\vspace{0.1cm}}
\newcommand{\eearms}{\end{eqnarray*}\vspace{0.1cm}}
\newcommand{\edmms}{\end{displaymath}\vspace{0.1cm}}

\newcommand{\bems}{\begin{multline*}}
\newcommand{\bem}{\begin{multline}}
\newcommand{\beams}{\vspace{0.1cm} \begin{eqnarray}}
\newcommand{\bearms}{\vspace{0.1cm} \begin{eqnarray*}}
\newcommand{\bdmms}{\vspace{0.1cm} \begin{displaymath}}

\newcommand{\be}{\begin{equation}}
\newcommand{\ee}{\end{equation}}

\newcommand{\bes}{\begin{equation*}}
\newcommand{\ees}{\end{equation*}}

\newcommand{\bea}{\begin{eqnarray}}
\newcommand{\eea}{\end{eqnarray}}

\newcommand{\bear}{\begin{eqnarray*}}
\newcommand{\eear}{\end{eqnarray*}}

\newcommand{\bdm}{\begin{displaymath}}
\newcommand{\edm}{\end{displaymath}}

\newcommand{\feop}{\hfill \rule{1.5mm}{1.5mm} \\}

\newcommand{\bit}{\begin{itemize}}
\newcommand{\eit}{\end{itemize}}

\newcommand{\ben}{\begin{enumerate}}
\newcommand{\een}{\end{enumerate}}

\newcommand{\bd}{\begin{document}}
\newcommand{\ed}{\end{document}}

\newcommand{\psp}{\hspace{0.6cm}}
\newcommand{\hspone}{\hspace{1cm}}
\newcommand{\hsptwo}{\hspace{2cm}}

\newcommand{\hspthree}{\hspace{3cm}}
\newcommand{\hspfour}{\hspace{4cm}}

\newcommand{\hspfive}{\hspace{5cm}}
\newcommand{\hspsix}{\hspace{6cm}}

\newcommand{\hspseven}{\hspace{7cm}}
\newcommand{\hspeight}{\hspace{8cm}}

\newcommand{\vspone}{\vspace{1cm}}
\newcommand{\vsptwo}{\vspace{2cm}}

\newcommand{\vspthree}{\vspace{3cm}}
\newcommand{\vspfour}{\vspace{4cm}}

\newcommand{\vspfive}{\vspace{5cm}}
\newcommand{\vspsix}{\vspace{6cm}}

\newcommand{\vspseven}{\vspace{7cm}}
\newcommand{\vspeight}{\vspace{8cm}}

\newcommand{\vspnine}{\vspace{9cm}}
\newcommand{\vspten}{\vspace{10cm}}

\newcommand{\vspeleven}{\vspace{11cm}}
\newcommand{\vsptwelve}{\vspace{12cm}}

\newcommand{\qett}{\mbox{$(q^{-1})$}}
\newcommand{\zett}{\mbox{$(z^{-1})$}}

\newcommand{\qo}{q^{-1}}
\newcommand{\zettu}{\mbox{$z^{-1}$}}

\newcommand{\linf}[1]{{#1}\rightarrow \infty}

\newcommand{\cf}[1]{(\ref{#1})}

\newcommand{\eiw} {\mathrm{e}^{\mathrm{i}\omega}}
\newcommand{\emiw}{\mathrm{e}^{-\mathrm{i}\omega}}

\newcommand{\tr}{\mbox{${\rm Tr}$}}
\newcommand{\var}{\mbox{${\rm Var}$}}
\newcommand{\diag}{\mbox{${\rm Diag}$}}
\newcommand{\cov}{\mbox{${\rm Cov}$}}
\newcommand{\Ex}{{\sf E}}

\newcommand{\erm}{\mathrm{e}}
\newcommand{\irm}{\mathrm{i}}
\newcommand{\Prj} {{\bf \Pi}^\bot}

\newcommand{\bA} {{\bf A}}
\newcommand{\bB} {{\bf B}}
\newcommand{\bC} {{\bf C}}
\newcommand{\bD} {{\bf D}}
\newcommand{\bE} {{\bf E}}
\newcommand{\bF} {{\bf F}}
\newcommand{\bG} {{\bf G}}
\newcommand{\bH} {{\bf H}}
\newcommand{\bI} {{\bf I}}
\newcommand{\bJ} {{\bf J}}
\newcommand{\bK} {{\bf K}}
\newcommand{\bL} {{\bf L}}
\newcommand{\bM} {{\bf M}}
\newcommand{\bN} {{\bf N}}
\newcommand{\bO} {{\bf O}}
\newcommand{\bP} {{\bf P}}
\newcommand{\bQ} {{\bf Q}}
\newcommand{\bR} {{\bf R}}
\newcommand{\bS} {{\bf S}}
\newcommand{\bT} {{\bf T}}
\newcommand{\bU} {{\bf U}}
\newcommand{\bV} {{\bf V}}
\newcommand{\bW} {{\bf W}}
\newcommand{\bY} {{\bf Y}}
\newcommand{\bX} {{\bf X}}
\newcommand{\bZ} {{\bf Z}}

\newcommand{\GLm} {{\bf \Lambda}}
\newcommand{\GSg} {{\bf \Sigma}}
\newcommand{\GPh} {{\bf \Phi}}
\newcommand{\GPs} {{\bf \Psi}}
\newcommand{\GPi} {{\bf \Pi}}
\newcommand{\GDl} {{\bf \Delta}}
\newcommand{\GGm} {{\bf \Gamma}}
\newcommand{\GUp} {{\bf \Upsilon}}
\newcommand{\GOm} {{\bf \Omega}}
\newcommand{\GTh} {{\bf \Theta}}
\newcommand{\GXi} {{\bf \Xi}}

\newcommand{\gal} {\boldsymbol{\alpha}}
\newcommand{\gbt} {\boldsymbol{\beta}}
\newcommand{\ggm} {\boldsymbol{\gamma}}
\newcommand{\gdl} {\boldsymbol{\delta}}
\newcommand{\gep} {\boldsymbol{\epsilon}}
\newcommand{\vep} {\boldsymbol{\varepsilon}}
\newcommand{\gzt} {\boldsymbol{\zeta}}
\newcommand{\get} {\boldsymbol{\eta}}
\newcommand{\gth} {\boldsymbol{\theta}}
\newcommand{\vth} {\boldsymbol{\vartheta}}
\newcommand{\git} {\boldsymbol{\iota}}
\newcommand{\gkp} {\boldsymbol{\kappa}}
\newcommand{\glm} {\boldsymbol{\lambda}}
\newcommand{\gmu} {\boldsymbol{\mu}}
\newcommand{\gnu} {\boldsymbol{\nu}}
\newcommand{\gxi} {\boldsymbol{\xi}}
\newcommand{\gpi} {\boldsymbol{\pi}}
\newcommand{\vpi} {\boldsymbol{\varpi}}
\newcommand{\grh} {\boldsymbol{\rho}}
\newcommand{\vrh} {\boldsymbol{\varrho}}
\newcommand{\gsg} {\boldsymbol{\sigma}}
\newcommand{\vsg} {\boldsymbol{\varsigma}}
\newcommand{\gtu} {\boldsymbol{\tau}}
\newcommand{\gup} {\boldsymbol{\upsilon}}
\newcommand{\gph} {\boldsymbol{\phi}}
\newcommand{\vph} {\boldsymbol{\varphi}}
\newcommand{\gch} {\boldsymbol{\chi}}
\newcommand{\gps} {\boldsymbol{\psi}}
\newcommand{\gom} {\boldsymbol{\omega}}

\newcommand{\Ba}  {{\bf a}}
\newcommand{\Bb}  {{\bf b}}
\newcommand{\Bc}  {{\bf c}}
\newcommand{\Bd}  {{\bf d}}
\newcommand{\Be}  {{\bf e}}
\newcommand{\Bf}  {{\bf f}}
\newcommand{\Bg}  {{\bf g}}
\newcommand{\Bh}  {{\bf h}}
\newcommand{\Bi}  {{\bf i}}
\newcommand{\Bj}  {{\bf j}}
\newcommand{\Bk}  {{\bf k}}
\newcommand{\Bl}  {{\bf l}}
\newcommand{\Bm}  {{\bf m}}
\newcommand{\Bn}  {{\bf n}}
\newcommand{\Bo}  {{\bf o}}
\newcommand{\Bp}  {{\bf p}}
\newcommand{\Bq}  {{\bf q}}
\newcommand{\Br}  {{\bf r}}
\newcommand{\Bs}  {{\bf s}}
\newcommand{\Bt}  {{\bf t}}
\newcommand{\Bu}  {{\bf u}}
\newcommand{\Bv}  {{\bf v}}
\newcommand{\Bw}  {{\bf w}}
\newcommand{\Bx}  {{\bf x}}
\newcommand{\By}  {{\bf y}}
\newcommand{\Bz}  {{\bf z}}

\newcommand{\za} {\boldsymbol{a}}
\newcommand{\zb} {\boldsymbol{b}}
\newcommand{\zc} {\boldsymbol{c}}
\newcommand{\zd} {\boldsymbol{d}}
\newcommand{\ze} {\boldsymbol{e}}
\newcommand{\zf} {\boldsymbol{f}}
\newcommand{\zg} {\boldsymbol{g}}
\newcommand{\zh} {\boldsymbol{h}}
\newcommand{\zi} {\boldsymbol{i}}
\newcommand{\zj} {\boldsymbol{j}}
\newcommand{\zk} {\boldsymbol{k}}
\newcommand{\zl} {\boldsymbol{\ell}}
\newcommand{\zm} {\boldsymbol{m}}
\newcommand{\zn} {\boldsymbol{n}}
\newcommand{\zo} {\boldsymbol{o}}
\newcommand{\zp} {\boldsymbol{p}}
\newcommand{\zq} {\boldsymbol{q}}
\newcommand{\zr} {\boldsymbol{r}}
\newcommand{\zs} {\boldsymbol{s}}
\newcommand{\zt} {\boldsymbol{t}}
\newcommand{\zu} {\boldsymbol{u}}
\newcommand{\zv} {\boldsymbol{v}}
\newcommand{\zw} {\boldsymbol{w}}
\newcommand{\zx} {\boldsymbol{x}}
\newcommand{\zy} {\boldsymbol{y}}
\newcommand{\zz} {\boldsymbol{z}}

\newcommand{\hide}[1] {}

\newtheorem{Thm}{Theorem}
\newtheorem{Pro}{Proposition}
\newtheorem{Defn}{Definition}
\newtheorem{Exm}{Example}
\newtheorem{Lem}{Lemma}
\newtheorem{Asm}{Assumption}
\newtheorem{Cor}{Corollary}
\newtheorem{Alg}{Algorithm}
\newtheorem{paRem}{Remark}
\newenvironment{Rem}{\begin{paRem}\rm }{\end{paRem}}

\newcommand{\ColvecTwo}[2]{\left[\ba{c} {#1} \\ {#2} \ea \right]}
\newcommand{\colvecTwo}[2]{\left[\ba{cc} {#1} & {#2} \ea \right]^\top}

\newcommand{\colvecThr}[2]{\left[\ba{ccc} {#1} & \cdots & {#2} \ea \right]^\top}
\newcommand{\ColvecThr}[2]{\left[\ba{c} {#1} \\ \vdots \\ {#2} \ea \right]}
\newcommand{\rowvecThr}[2]{\left[\ba{ccc} {#1} & \cdots & {#2} \ea \right]}
\newcommand{\MatThrTwo}[4]{\left[\ba{ccc} {#1}  & {#3} \\ \vdots & \vdots \\ {#2} & {#4} \ea \right]}

\newcommand{\md}{\mathrm{d}}
\newtheorem{lem}{Lemma}
\newcommand\myeq{\mathrel{\stackrel{\makebox[0pt]{\mbox{\normalfont\tiny def}}}{=}}}

\author{ 
}

\begin{document}

\title{Multiuser Detection for Random Access Bandwidth Request in WiMAX}

\author{\IEEEauthorblockN{Md Mashud Hyder}\\
 \IEEEauthorblockA{School of Electrical Engineering and Computer
Science\\
 The University of Newcastle, NSW 2308, Australia}%
}
\maketitle
\begin{abstract}
Random access is a multiple access communication protocol where the users simultaneously communicate with a base station (BS) in an uncoordinated fashion. In this work, we consider the problem of multiuser detection in a random access bandwidth request context. We propose an enhanced random access scheme where the fixed/low-mobility M2M devices pre-equalize their random access codes using the estimated frequency response of the slowly-varying wireless channel. Consequently, we have developed two different multiuser detection algorithms. The first algorithm works in a greedy fashion where it performs cross-correlation of the received signal with a set of decoder sequences and detects active users based on the correlation output. We derive the condition under which the algorithm can detect a given number of active users with high probability. Subsequently, we demonstrate an efficient decoder design procedure which enhances the user detection performance. A basis mismatched sparse recovery technique has been applied in the second algorithm which exploit an inherent structure of the random access protocol. The performance of the proposed schemes is demonstrated in a WiMAX network environment.

\end{abstract}

\section{Introduction}
The orthogonal frequency-division multiple access (OFDMA) scheme has been adopted by the IEEE 802.16 WiMAX standards. The future wireless communication network has to support a large number of fixed/low-mobility machine-to-machine (M2M) devices that will transmit bursty, small data packets (e.g. meter readings, sensor reports etc.) under a valid security association with the network \cite{2010a}. This corresponds to a heavily uplink-based traffic model following a Poisson distribution. Random access based bandwidth request (BR) is preferable for such traffic that exploits the benefits of statistical multiplexing to support a large number of devices with a fixed overhead. In the IEEE WiMAX standards, the BR procedure starts with the allocation of a predefined set of subcarriers by the BS in a specific time slots. The subscriber stations (SSs) which wish to request for bandwidth to the BS can take the opportunity by modulating a randomly selected code onto the allocated subcarriers. At the receiver end, the BS is required to detect the multiple active subscribers. Once the BR subscribers are resolved, the BS will allocate bandwidth for the corresponding subscribers. However, the process will be efficient when the base station can separate the BR subscribers successfully.

The idea of random access channel is related to the concept of multiple access channel (MAC) in network information theory \cite{mac}. The concept of MAC has been applied successfully in CDMA systems \cite{cdma}. However, the application of classic MAC channel analysis in the present scenario is not straightforward (see \cite{mud8} and references therein). The problem of multiuser detection (MUD) in a random access on-off channel has been studied in \cite{mud8,mud9,mud4}. The work in \cite{mud8} solves the MUD problem using a sparse signal recovery framework and derives necessary conditions under which the orthogonal matching pursuit and Lasso algorithms can detect active users successfully. However, the proposed method assumes perfect synchronization among all users, which is difficult to guarantee in practice. A modified sparse representation framework has been proposed in \cite{mud9} where the above assumption has been relaxed. A reduced dimension MUD method has been proposed in \cite{mud4}. It has been shown that the computational complexity of the reduced dimension MUD is lower than conventional correlation based MUD. However, all the methods assume that the transmitted signal is subject to flat fading channel. In a metropolitan wireless network the communication channel generally subject to frequency selective fading and hence the above assumption may not hold in practice. Furthermore, the sparse representation of the received signal is not straightforward for selective fading channel.

Multiuser detection in IEEE 802.16 based BR has been studied in \cite{krinock2001comments,6327306,6461488,6554572}. A simple solution to enhance MUD is to increase the number of BR channels to accommodate more users/devices. However, under the existing
schemes, only a handful of codes can be detected reliably per channel
per frame in presence of multiple access interference (MAI) from different
codes, and random noise and frequency-selective fading in the multipath
wireless channel \cite{krinock2001comments}. Therefore, a lot of
random access channels are required, which would substantially reduce the payload
capacity of the overall system. Hence, random access has been considered
as one of the key bottlenecks by the IEEE 802.16p working group on
M2M communications. To overcome this limitation, the recent IEEE 802.16p
amendment has proposed several solutions, mostly based on access control
over the MAC layer \cite{6327306}. On the other hand, a number of
works have already appeared in the literature concerning this problem,
e.g. \cite{6461488} and \cite{6554572}. However, these schemes require
significant modification to the existing standards and are not suitable
for a network supporting both M2M and non-M2M traffic.

In this work we propose an enhanced random access MUD in M2M communication environment. The MUD problem has been resolved by two different algorithms. The contributions of the work can be summarized as follows:
\begin{enumerate}
\item We propose a pre-equalized random access scheme. We assume that the BR subscribers only have an approximate knowledge about their channel frequency response. The BR subscribers pre-equalize the random access codes by using the estimated channel frequency response and transmit to the BS. The imperfect knowledge of channel frequency response results an additional noise term in the received signal at the BS. We analyze the statistical property of the noise term and develop a suitable data model of the received signal for MUD.
\item We apply a very simple correlation based approach called CMUD for resolving the MUD problem where the received signal is correlated with a set of decoder sequences. We develop the necessary condition under which the CMUD can resolve a given number of active users from the received signal.
\item The necessary condition will show that one can enhance the performance of CMUD by controlling some properties of the decoder sequences. We then propose an algorithm to design efficient decoder sequences.  
\item The computational complexity of CMUD is very low, but it cannot detect a large number of active users. To enhance the user detection performance, we resolve the MUD problem by using a basis mismatched sparse signal recovery algorithm. 
We exploit some inherent properties of MUD and formulate an optimization problem which can enhance the performance of the underlying basis mismatched sparse recovery algorithm. However, the optimization problem is non-convex in general. We apply a Lagrangian Dual Relaxation method to solve the optimization problem.
\end{enumerate}

{\bf Notations:} Superscript $\top$ denotes matrix transpose. $\Ex(\zx)$ denotes expected value of $\zx$. For a complex number $\zx$, its real and imaginary parts will be denoted by $[\zx]_r$ and $[\zx]_i$ respectively. A component of the matrix $\mathbf{C}$ at its $l$-th row and $j$-th column will be indicated by $\bC_{l,j}$ and the $j$\textsuperscript{th} column of $\mathbf{C}$ will be represented by $\bC_j$. The cardinality of a set $\mathbb{T}$ will be denoted by $\#\mathbb{T}$. ${\bf 1}_L$ denotes a vector of length $L$ whose all components are one. The $\ell_p$ norm of a vector is defined as $\|\zx\|_p=\left(\sum_t|\zx_t|^p\right)^{1/p}$. diag$(\zx)$ refers to a diagonal matrix with vector $\zx$ on its diagonal.

\section{Data Model and Problem Statement}
\subsection{General System Model}
Consider a single-cell WiMAX network with time division duplex (TDD)
OFDMA physical layer. A BR channel is comprised of $L$ randomly
chosen subcarriers over one OFDM uplink symbol. Suppose the indices of BR subcarriers are $\{j_m : m = 1, 2, \ldots,L\}$. 
When a subscriber station (SS) intends to send a BR to the base station then it
selects an available uplink BR slot and sends a BR packet to the BS.
The packet consists of a BR code, which is a $L$-bit pseudo
random binary sequence (PRBS) chosen with equal probability from a
bank of $K$ codes \cite{5062485}, where $L<K$. For rest of
the sequel, we denote the BR code-matrix as $\mathbf{C}\in\mathbb{R}{}^{L\times K}$ where every column of $\bC$ represents an independent code. Each $\bC_{l,j}$ is modulated by binary phase shift keying (BPSK) i.e. $\bC_{l,j}\in\{-1,+1\}$. The code-matrix is known to BS and every subscriber. 

\subsection{Pre-Equalization}
In this proposed random access model, the SS does not send the random access code directly to the BS, instead it transmits a pre-equalized version of the code. According to the IEEE 802.16 standards, the first OFDM symbol of each WiMAX frame is a preamble transmitted
by the BS, where the subcarriers are BPSK modulated with a boosted
pilot sequence \cite{5062485}. Typically, the SSs use this information
to estimate the channel frequency response (CFR) for the OFDM demodulation
process. A SS can pre-equalize its BR code using this estimated CFR
exploiting the channel reciprocity of the TDD system \cite{851577}.
A number of pre-equalization techniques are available. For more details,
please refer to \cite{fazel2008multi}.

Consider a time instant when $M$ number of SSs are simultaneously
contending on the same BR channel. Let, the $m$\textsuperscript{th}
SS selects the $k_{m}$\textsuperscript{th} column of the code matrix
$\mathbf{C}\in\mathbb{R}{}^{L\times K}$, where $m=1,2,...,M$. Considering
zero-forcing (ZF) pre-equalization \cite{fazel2008multi}, the transmitted
code over the $j_l$\textsuperscript{th} subcarrier from the $m$\textsuperscript{th}
SS be

\begin{equation}
x_{l,m}=\frac{\mathbf{C}_{l,k_{m}}}{\hat{h}_{l,m}}; \ \textrm{for\;\ensuremath{\forall}\;}l\in\{1,2,...,L\},
\end{equation}

\noindent where $\hat{h}_{l,m}$ is the pilot-aided CFR of  $j_l$\textsuperscript{th} subcarrier estimated
by the $m$\textsuperscript{th} SS. To be more precise, $\hat{h}_{l,m}$
can be expressed as 
\begin{equation}
\hat{h}_{l,m}=h_{l,m}+e_{l,m},\label{eq:Pilot_CFR}
\end{equation}
where $h_{l,m}$ is the actual frequency response of the $j_l$\textsuperscript{th}
subcarrier and $e_{l,m}\sim\mathcal{C\mathcal{\mathcal{N}}}(0,\sigma_{e,m}^{2})$
is a zero-mean complex Gaussian noise with variance $\sigma_{e,m}^{2}$. 

In the BS, after down-conversion to baseband and OFDM demodulation,
the received signal from the $m$\textsuperscript{th} SS over the
$j_l$\textsuperscript{th} subcarrier be

\begin{equation}
y_{l,m}=x_{l,m}\bar{h}_{l,m}=\mathbf{C}_{l,k_{m}}\frac{\bar{h}_{l,m}}{\hat{h}_{l,m}}; \ \textrm{for\;\ensuremath{\forall}\;}l\in\{1,2,...,L\},
\end{equation}

\noindent where $\bar{h}_{l,m}$ is the effective channel experienced
by the BS due to transmission form $m$-th SS over the $j_l$-th subcarrier. 
Note that, although the position of the BS and the SS will remain almost fixed for
stationary/slow-moving M2M devices, the channel is continually affected by the
movement of the external scatterers in the surrounding environment
\cite{andersen2009doppler}. Consequently, $\bar{h}_{l,m}$ will differ
from $h_{l,m}$. To generalize, the effective channel can be
modelled as \cite{5683407}

\noindent 
\begin{equation}
\bar{h}_{l,m}=\alpha h_{l,m}+\eta_{l,m},
\end{equation}
where $\eta_{l,m}\sim\mathcal{C\mathcal{\mathcal{N}}}(0,\sigma_{\eta,m}^{2})$
is a zero-mean complex Gaussian noise and $\alpha$ is some deterministic
complex valued constant. 

Considering the above phenomena, the combined received signal at the BS from
all $M$ stations over the $j_l$\textsuperscript{th} subcarrier will
be
\begin{equation}
y_{l}=\sum_{m=1}^{M}\left\{ \mathbf{C}_{l,k_{m}}\frac{\alpha h_{l,m}+\eta_{l,m}}{h_{l,m}+e_{l,m}}\right\} +\vartheta_{l},\label{eq:Received_Signal}
\end{equation}
where $\vartheta_{l}\sim\mathcal{C\mathcal{\mathcal{N}}}(0,\sigma_{\vartheta}^{2})$
is a zero-mean complex Gaussian noise. 

Equation \eqref{eq:Received_Signal} can be represented as
\begin{equation}
y_{l}=\sum_{m=1}^{M}\left\{ \mathbf{C}_{l,k_{m}}-\lambda_{l,k_m}\mathbf{C}_{l,k_{m}}\right\} +\vartheta_{l},\label{eq:sum2}
\end{equation}
where $\lambda_{l,k_m}$ is a ratio of complex variables, i.e.
\begin{equation}
\lambda_{l,k_m}=\frac{(1-\alpha)h_{l,m}+e_{l,m}-\eta_{l,m}}{h_{l,m}+e_{l,m}}.\label{eq:lambda}
\end{equation}
Let us construct
\begin{equation}
\zy=[y_1,y_2,...,y_L]^{\top}.\label{eq:y}
\end{equation}

\subsection{Formal Problem Statement}
Let us define the active users set as 
\begin{align}
\mathcal{S}=\{k_m:m=1,2,\cdots M\}
\end{align}
Our goal is to detect the set of active codes, i.e., $\mathcal{S}$ from the received noisy data $\zy$.

\subsection{Sparse Representation}\label{sec:sp}
Typically, at a particular BR opportunity under a BS, the total number of BR terminals $M \ll K$. This condition should be met by any random access based communication network. Otherwise, by using probability theory, it can be shown that two or more BR terminals will collide by selecting the same BR code with high probability. It is well known that if multiple users transmit same code then the BS cannot separate the corresponding users. To maintain $M \ll K$, different approaches have been considered by the Third Generation Partnership Project \cite{lte6}. For example, the collision resolution algorithms such as random backoff can limit the value of $M$ \cite{lte7}. In the following, we shall assume that any BR code will be used by at most one BR terminal at a particular BR opportunity \footnote{In practice, multiple BR terminals may collied by selecting same BR code, however, this event will occur with very low probability. Furthermore, we observed using simulations that this rare event does not affect the performance of the proposed algorithms significantly.}. Let $\mathring{\zx}\in \mathbb{R}^K$ be a vector such that its $i$-th component will be $1$ only if $i\in\mathcal{S}$, and zero otherwise.  Using the vector we can represent the data model in \eqref{eq:y} as
\begin{align}\label{eq:dat2}
\zy=\bC\mathring{\zx}  -\zu+\vartheta 
\end{align}
where $\zu=\bQ\mathring{\zx}$, and 
\begin{align}\label{eq:Q}
\bQ=\left[\begin{array}{cccc}
\bC_{1,1}\lambda_{1,1} & \bC_{1,2}\lambda_{1,2} & \cdots &\bC_{1,K}\lambda_{1,K}\\
\bC_{2,1}\lambda_{2,1} & \bC_{2,2}\lambda_{2,2} & \cdots &\bC_{2,K}\lambda_{2,K}\\
\vdots & \vdots &\vdots &\vdots\\
\bC_{L,1}\lambda_{L,1} & \bC_{L,2}\lambda_{L,2} & \cdots &\bC_{L,K}\lambda_{L,K}
\end{array}\right],
\end{align}
with $\bQ_i=0$ if $i\not\in\mathcal{S}$. Since $M\ll K$ by assumption, the vector $\mathring{\zx}$ is sparse. Thus, MUD can be seen as a problem of estimating the sparse vector $\mathring{\zx}$ from the received data $\zy$. Now suppose that we have some prior knowledge about the number of active BR users i.e., $M$. Then we can postulate that $\mathring{\zx}_i$ has a probability density function
\begin{align}\label{eq:xpdf}
p_{x}(\mathring{\zx}_i)=(1-\epsilon)\delta_0(\mathring{\zx}_i)+\epsilon \delta_1(\mathring{\zx}_i)
\end{align}
where $\epsilon=\frac{M}{K}$, $\delta_1(x)=\delta_0(x-1)$ and $\delta_0(x)$ is a Dirac delta function.

\section{Correlation based code detector}\label{sec:corr}
In this section, we apply a simple correlation based algorithm to detect active BR codes. 
In detecting active codes, we take real part of $\zy$ in \eqref{eq:dat2} because imaginary part contains only noise and interference. In the following, we shall study some statistical properties of $\lambda$ which will be helpful for developing the code detection algorithm. 
\subsection{Statistical Properties of $\lambda$}\label{sec:lam}
In general, the components of CFR are modelled as zero mean complex Gaussian random variable \cite{5683407}, i.e, $h_{l,m}\sim\mathcal{C\mathcal{\mathcal{N}}}(0,\sigma_{h,m}^{2})$. Hence, the channel
power can be approximated by its variance. The variance of CFR, i.e. $\sigma_{h,m}^{2}$
may not same for every $m=1,2,\cdots M$. However, the BS always
equalizes the channel power of the active users through the ranging
procedure. In effect, the variances of CFR of all users must
remain in a known interval due to the ranging process. Consequently,
we can use the value of average channel power as an estimate of $\sigma_{h,m}^{2}$
for all $m$. 
Similarly, BS can estimate \textbf{$\sigma_{e,m}^{2}$ }and
$\sigma_{\eta,m}^{2}$ by using pilot-aided synchronization procedure. Thereby, we shall use the following assumption.
\begin{Asm}\label{asm1}
$\sigma^2_{h,i}=\sigma^2_{h},\sigma^2_{e,i}=\sigma^2_{e},\sigma^2_{\eta,i}=\sigma^2_{\eta}; \forall i\in\{1,2,\cdots M\}$. 
\end{Asm}

According to \eqref{eq:lambda},
$\lambda_{l,m}$ is a ratio of two complex quantities. By applying the concept of \cite{5683407} and using Assumption-\ref{asm1}, every $\lambda_{l,m}$ can be modelled using a complex random
variable $\lambda=[\lambda]_{r}+\textrm{i}[\lambda]_{i}$ with probability density function:
\begin{align}
 & f([\lambda]_{r},[\lambda]_{i})=\nonumber \\
 & \frac{(1-|\rho|^{2})\sigma_{u}^{2}\sigma_{v}^{2}}{\pi\left(\sigma_{v}^{2}|\lambda|^2+\sigma_{u}^{2}-2\rho_{r}[\lambda]_{r}\sigma_{u}\sigma_{v}+2\rho_{i}[\lambda]_{i}\sigma_{u}\sigma_{v}\right)^{2}}\label{eq:pdf}
\end{align}
where $\sigma_{u}^{2}=|1-\alpha|^{2}\sigma_{h}^{2}+\sigma_{e}^{2}+\sigma_{\eta}^{2}$,
$\sigma_{v}^{2}=\sigma_{h}^{2}+\sigma_{e}^{2}$, and $\rho=(1-\alpha)\sigma_{h}^{2}+\sigma_{e}^{2}$. Furthermore, $\{\lambda_{l,m}\}_{l,m}$ are independent from each other. 
In Appendix-A, we derive the mathematical expectation and variance
of $[\lambda]_{r}$. In the following we denote them by $\mu_{r}$ and $\sigma_{r}^{2}$ respectively.

Since $\bC_{\ell,j}$ are real valued, the expected value of $[\zu_j]_r$ in \eqref{eq:dat2} with respect to $\mathring{\zx}$ and $\lambda$ is
\begin{align}\label{eq:mlam}
\Ex([\zu_j]_r)&=\Ex\left(\sum_{\ell=1}^K{\bC_{j,\ell}[\lambda_{j,\ell}]_r\mathring{\zx}_{\ell}}\right)\nonumber\\
&=\sum_{\ell=1}^K\left[\left(\bC_{j,\ell}\Ex([\lambda]_r)\right)p_x(\mathring{\zx}_{\ell}=1)\right]\nonumber\\
&=\epsilon \ \mu_r \ \sum_{\ell=1}^K\bC_{j,\ell}
\end{align}
By using a similar procedure, it can be verified that the covariance matrix of $[\zu]_r$ is
\begin{align}
\mathrm{Cov}([\zu]_r)=\epsilon \ \sigma_r^2\left[\begin{array}{ccc}
\|\bC(1,:)\|_2^2  & \cdots &0\\
0  & \cdots &0\\
\vdots  &\vdots &\vdots\\
0 &  \cdots &\|\bC(K,:)\|_2^2
\end{array}\right]\nonumber
\end{align}
where $\bC(\ell,:)$ denotes $\ell$-th row of $\bC$. In the present scenario $\bC_{j,\ell}\in\{+1,-1\}$, hence
\begin{align}\label{eq:vlam}
\mathrm{Cov}([\zu]_r)=M\sigma_r^2\bI.
\end{align}

\subsection{Correlation based multiuser detection (CMUD) algorithm}
The CMUD algorithm is summarized in Table-\ref{tab:algo}. It starts with an empty set $\mathbb{T}$ as an initial estimate of indices of active codes and residual vector $\zz^{(0)}=\zy$. To detect the presence of an active BR code, the code detector correlates
the residual with a decoder matrix $\mathbf{D}\in\mathbb{R}{}^{L\times K}$.
In Step 2 of $j$-th iteration, the algorithm takes every column of $\mathbf{D}$ and computes its
cross-product with the residual $\zz^{(j-1)}$. It then construct a set $\mathbb{I}$ such that
\begin{align}
\mathbb{I}=\{\ell\in\{1,2,\cdots K\}: |[\bD_{\ell}^{\top}\zz^{(j-1)}]_r|>\kappa \ \}\nonumber
\end{align}
where $\kappa$ is a predefined threshold. In Step 3, the set $\mathbb{I}$ is added to the active code set i.e., $\mathbb{T}=\mathbb{T}\cup\mathbb{I}$. However, if it is found that $\mathbb{I}$ is empty then we assume all active codes in $\zy$ has been detected by the algorithm and hence terminate to Step 6. In Step 4, the residual $\zz^{(j)}$ is updated by subtracting the selected active codes from $\zz^{(j-1)}$. The residual $\zz^{(j)}$ represents the part of active codes $\{\bC_{\ell}\}_{\ell\in\mathcal{S}}$ that has not been detected yet along with noise. In Step 5, we update $\kappa$ to a new value and repeat Steps 1-5. The algorithm needs a decoder matrix $\bD$ and threshold $\kappa$ as its input. In the following section, we shall demonstrate some procedures for choosing those parameters. Furthermore, the algorithm needs an estimate of $M$. We shall describe a simple procedure in Section-\ref{sec:muser} to obtain a rough estimate of the parameter. When $\bD=\bC$, the CMUD will be closely related to the reduced dimension decision-feedback (RDDFt) algorithm \cite{mud4}. However, we shall show that an appropriate choice of $\bD$ can increase the performance of CMUD significantly. The computational complexity of CMUD is very small. At every iteration, the major complexity involves in computing $[\bD^{\top}\zr^{(j-1)}]_r$, which requires $LK$ flops.

\begin{table}[!t]
\renewcommand{\arraystretch}{1.3}
\centering \caption{Correlation based multiuser detection (CMUD)}\label{tab:algo}
\begin{tabular}{l}
 \hline
 {\bf Input:} Code matrix $\bC$, a real valued decoder matrix $\bD\in\mathbb{R}^{L\times K}$, \\
 \ \ \ \ \ \ \ date vector $\zy$, threshold $\kappa$ and an approximation of $M$, i.e., $M_0$.\\
{\bf Initialization:} Set $\mathbb{T}=\emptyset, \zz^{(0)}=\zy, j=0$.\\
{\bf Loop: } for $j=1,2\cdots M_0$\\
\ \ \ \ 1. Set $\mathbb{I}=\emptyset$.\\
\ \ \ \ 2. Construct $\mathbb{I}=\{\ell\in\{1,2,\cdots K\}: |[\bD_{\ell}^{\top}\zz^{(j-1)}]_r|>\kappa \ \}$.\\
\ \ \  \ 3. If $\mathbb{I}\not=\emptyset$: Set $\mathbb{T}=\mathbb{T}\cup\mathbb{I}$.\\
\ \ \ \ \ \  \ Else: Go to Step 6.\\
\ \ \ \ 4. Update $\zz^{(j)}=\zz^{(j-1)}-\sum_{\ell\in\mathbb{I}}\bC_{\ell}$.\\
\ \ \ \ 5. Update $\kappa$.\\
{\bf End Loop} \\
6. {\bf Output:} Indices of active codes $\hat{\mathbb{T}}=\mathbb{T}$. \\
\hline
\end{tabular}
\end{table}

\subsection{Performance Analysis}
In this section, we develop the conditions under which the CMUD will successfully detect all active codes. Our performance measure is based on the probability of code detection error by CMUD. In particular, we define the probability of code detection error as
\begin{align}\label{eq:pe}
P_e=Pr\{\hat{\mathbb{T}}\not=\mathcal{S}\}.
\end{align}
Let us define
\begin{align}\label{eq:d1}
\alpha&=\max_{\ell}|\sum_{j=1}^K{\bD_{\ell}^{\top}\bC_{j}}|\\
\beta&=\max_{\ell}\{\max_{j; j\not=\ell}|\bD_{\ell}^{\top}\bC_{j}|\}\\
\label{eq:d3}\gamma&=\max_{\ell}\|\bD_{\ell}\|_2.
\end{align}

\begin{Lem}\label{lem0}
Let $\zy=\sum_{\ell\in\mathcal{S}}\bC_{\ell}-\zu+\vartheta$ where $\#\mathcal{S}=M$ and the random vector $\vartheta$ has zero mean complex Gaussian distribution with covariance $\sigma_{\vartheta}^2\bI$ and the random vector $[\zu]_r$ is defined as in \eqref{eq:dat2}. Let $\bD\in\mathbb{R}^{L\times K}$ be a decoder matrix such that $\bD_{\ell}^{\top}\bC_{\ell}=1;\forall \ell$. Set
\begin{align}\label{eq:tau}
\tau=\epsilon\mu_r\alpha+\gamma\sqrt{2(1+\nu)\log K}\sqrt{M_0\sigma_r^2+\sigma_{\vartheta}^2/2}
\end{align} 
for some given $\nu>0$ and $M_0\ge M$. Assume that the decoder matrix $\bD$  satisfies the following condition:
\begin{align}\label{eq:d}
\tau+M_0\beta<0.5.
\end{align}
If we choose a threshold $\kappa$ such that
\begin{align}\label{eq:k}
\tau+M_0\beta< \kappa< 1-M_0\beta-\tau
\end{align}
then the probability of code detection error \eqref{eq:pe} by CMUD will be upper bounded by 
\begin{align}
P_e\le (\pi(1+\nu)\log K)^{-1/2}K^{-\nu}.
\end{align}
\end{Lem}
{\bf Proof:} See Appendix-B. \feop

\subsection{Decoder Design}
Lemma-\ref{lem0} states that for given values of $M$ and noise variances, the CMUD will detect all active codes efficiently if the matrix $\bD$ satisfies the following two conditions:
\begin{align}\label{eq:c1}
\epsilon\mu_r\alpha+\gamma\Upsilon+M_0\beta<0.5.\\
\label{eq:c2}\bD_{\ell}^{\top}\bC_{\ell}=1; \ \ \forall \ell.
\end{align}
where we denote $\Upsilon=\sqrt{2(1+\nu)\log K}\sqrt{M_0\sigma_r^2+\sigma_{\vartheta}^2/2}$.
Hence, to obtain an optimum decoder, we have to design $\bD$ that minimizes the left side of \eqref{eq:c1}. In Section-\ref{sec:dec1} we shall demonstrate a procedure of decoder design based on the objective. As will be seen later, the procedure requires to solve a high dimensional optimization problem. Note that the base station needs not to redesign $\bD$ at every BR request interval, instead it will be redesigned only when the values of $M$ and noise variances change significantly. Thereby, the proposed scheme will be compliant with the IEEE 802.16 standards. Nevertheless, the base station always seeks low complex algorithm. A low complex decoder design procedure will be proposed in Section-\ref{sec:dec2} which adopts a popular approach called minimum mean square error (MMSE) decoder \cite{mud7}. The MMSE decoder will not minimize the left side of \eqref{eq:c1} directly, however, exhibits moderate number of BR code detection performance.

\subsubsection{Dectoder-I}\label{sec:dec1}
Let  $\hat{\bC}(\ell)$ be a matrix constructed from $\bC$ by retaining all its columns except the $\ell$-th column. To obtain an optimum decoder based on Lemma-\ref{lem0}, we need to solve the following optimization problem:
 \begin{align}\label{eq:opt4}
\{\bD_*,\hat{\alpha}_*,\hat{\beta}_*,\hat{\gamma}_*\}=&\arg \min_{\bD,\hat{\alpha},\hat{\beta},\hat{\gamma}} \ \epsilon\mu_r\hat{\alpha}+M_0\hat{\beta}+\Upsilon \hat{\gamma}\\
\mathrm{subject \ to,}& \ \bD_{\ell}^{\top} \bC_{\ell}=1, \ \mathrm{for} \ \ell=1,2,\cdots K \nonumber\\
-\hat{\alpha} \le &[\sum_{j=1}^K\bC_j]^{\top}\bD_{\ell}\le \hat{\alpha}, \ \mathrm{for} \ \ell=1,\cdots K \nonumber\\
-\hat{\beta}\bf{1}_{K-1}\le&[\hat{\bC}(\ell)]^{\top}\bD_{\ell}\le\hat{\beta}\bf{1}_{K-1}, \ \mathrm{for} \ \ell=1,\cdots K  \nonumber\\
&\|\bD_{\ell}\|_2\le \hat{\gamma}, \ \mathrm{for} \ \ell=1,2,\cdots K.\nonumber
\end{align}
The optimization is convex and can be solved efficiently by using a Primal-Dual algorithm \cite{co}. However, the optimization needs to solve for $L.K+3$ number of variables. In the following, we propose a low complex decoder design procedure.

\subsubsection{Dectoder-II}\label{sec:dec2}
The decoder design strategy is based on the MMSE criterion \cite{mud7}. In particular, for every $\ell\in\{1,2,\cdots K\}$, we design $\bD_{\ell}\in\mathbb{R}^{L}$ that minimizes $\Ex(\|\mathring{\zx}_{\ell}-\bD_{\ell}^{\top}[\zy]_r\|_2^2)$ along with the constraint $\bD_{\ell}^{\top}\bC_{\ell}=1$. Here, the expectation is with respect to $\mathring{\zx}$, $\lambda$ and noise vector $\vartheta$. Let us define the index set $\mathbb{U}:=\{1,2,\cdots K\}$. By using \eqref{eq:dat2} and the fact that $\bD_{\ell}^{\top}\bC_{\ell}=1$, we see that
\begin{align}
\bD_{\ell}^{\top}\zy=\mathring{\zx}_{\ell}+\sum_{j\in\mathbb{U}\setminus \ell}\bD_{\ell}^{\top}\bC_{j}\mathring{\zx}_j-\sum_{j\in\mathbb{U}}\bD_{\ell}^{\top}\bQ_j\zx_j+\bD_{\ell}^{\top}\vartheta
\end{align}
Note that the value of $\epsilon$ in \eqref{eq:xpdf} is small. Hence, neglecting the values of $\Ex\left(\mathring{\zx}_j\bD_{\ell}^{\top}\bC_j\bQ_k^{\top}\bD_{\ell}\mathring{\zx}_k\right);\forall j\not=k$ and $\Ex\left(\mathring{\zx}_j\bD_{\ell}^{\top}\bQ_j\bQ_k^{\top}\bD_{\ell}\mathring{\zx}_k\right);\forall j\not=k$, and using a similar procedure of \eqref{eq:mlam}, we obtain that
\begin{align}
\Ex&(\|\mathring{\zx}_{\ell}-\bD_{\ell}[\zy]_r\|_2^2)\\
&=\epsilon\sum_{j\in\mathbb{U}\setminus \ell}\|\bD_{\ell}^{\top}\bC_{j}\|_2^2+\epsilon\sum_{j\in\mathbb{U}}\Ex(\|\bD_{\ell}^{\top}[\bQ_{j}]_r\|_2^2)\nonumber\\
& \ \ \ -2\epsilon\sum_{j\in\mathbb{U}\setminus \ell}\Ex(\bD_{\ell}^{\top}\bC_j[\bQ_j]_r^{\top}\bD_{\ell})+\|\bD_{\ell}\|_2^2\sigma_{\vartheta}^2/2\nonumber\\
&\le \epsilon\sum_{j\in\mathbb{U}}\|\bD_{\ell}^{\top}\bC_{j}\|_2^2+\epsilon\sum_{j\in\mathbb{U}}\|\bD_{\ell}^{\top}\bC_{j}\|_2^2\Ex(\lambda^2)\nonumber\\
& \ \ \ -2\epsilon\sum_{j\in\mathbb{U}\setminus \ell}\|\bD_{\ell}^{\top}\bC_{j}\|_2^2\Ex(\lambda)+\|\bD_{\ell}\|_2^2\sigma_{\vartheta}^2/2\nonumber\\
\label{eq:r} &\approx\delta\bD_{\ell}^{\top}\bR\bD_{\ell}+\|\bD_{\ell}\|_2^2\sigma_{\vartheta}^2/2
\end{align}  
where $\delta= \epsilon(1+\Ex([\lambda]_r^2)-2\mu_r)$ and $\bR=\sum_{j\in \mathbb{U}}\bC_j\bC_j^{\top}$. Now we have to minimize \eqref{eq:r} with respect to $\bD_{\ell}$ along with the constraint $\bD_{\ell}^{\top}\bC_{\ell}=1$. Using the procedure of Lagrangian multiplier, it can be verified that the optimization has a closed form of solution:
\begin{align}\label{eq:mmse}
\bD_{\ell}=\frac{(\delta\bR+\bI\sigma_{\vartheta}^2/2)^{-1}\bC_{\ell}}{\bC_{\ell}^{\top}(\delta\bR+\bI\sigma_{\vartheta}^2/2)^{-1}\bC_{\ell}}.
\end{align}
Note that whenever the number of active users or channel noise variance changes significantly, we have to compute the inverse term $(\delta\bR+\bI\sigma_{\vartheta}^2/2)^{-1}$ to obtain an updated $\bD$, which is computationally expensive. We can reduce the computational complexity by using the following procedure. Since $L<K$, the matrix $\bR$ can be shown to be positive definite (see \cite{rad}). Hence, the eigenvalue decomposition of $\bR$ has the form: $\bR=\bU\Lambda\bU^{\top}$ where $\bU^{\top}\bU=\bI$ and $\Lambda$ is a diagonal matrix whose $(k,k)$-th entry is the $k$-th eigenvalue of $\bR$. Denote $\zv=\bU^{\top}\bC_{\ell}$. Then it can be verified that
\begin{align}\label{eq:dec1}
\bD_{\ell}=\frac{U(\delta\Lambda+\bI\sigma_{\vartheta}^2/2)^{-1}\zv}{\sum_{j=1}^L\frac{\zv_j^2}{\delta\Lambda_{j,j}+\sigma_{\vartheta}^2/2}}
\end{align}
Since $(\delta\Lambda+\bI\sigma_{\vartheta}^2/2)$ is a diagonal matrix, its inverse can be computed easily. Furthermore, for a given $\bC$ the matrix $\bR$ is fixed, hence we can compute the eigenvalue decomposition in priori. Note that the minimization of \eqref{eq:r} will minimize the values of $\alpha$ and $\gamma$ in \eqref{eq:c1}. However, it does not consider minimizing $\beta$.

\subsection{Estimate of total number of active users}\label{sec:muser}
 The CMUD algorithm in Table-\ref{tab:algo} requires an approximation of the total number
of active users $M$ which is unknown in practice. 
In the following, we demonstrate an approximate procedure to estimate $M$. 
The energy received
at BS due to real parts of $\mathbf{y}$ be (using \eqref{eq:sum2})
\begin{align}
\mathcal{G} & =[\mathbf{y}]_r^{T}[\mathbf{y}]_r=\sum_{\ell=1}^{L}\left[\sum_{m=1}^{M}\left\{ \mathbf{C}_{\ell,k_{m}}-[\lambda_{\ell,m}]_r\mathbf{C}_{\ell,k_{m}}\right\} +[v_{\ell}]_r\right]^{2}
\end{align}
Taking expectation of $\mathcal{G}$ with respect to $\vartheta$ and $\lambda$, we see
\begin{align}\label{eq:eg}
\Ex(\mathcal{G})&=\Ex\Bigg(\sum_{\ell=1}^L\bigg\{\sum_{m=1}^M\left\{\bC_{\ell,k_{m}}-[\lambda_{\ell,m}]_r\bC_{\ell,k_{m}}\right\}^2 \nonumber\\
+2\sum_{m=1}^M&\sum_{\substack{j=1\\j\not=m}}^M\{\bC_{\ell,k_{m}}-[\lambda_{\ell,m}]_r\bC_{\ell,k_{m}}\} \{\bC_{\ell,k_{j}}-[\lambda_{\ell,j}]_r\bC_{\ell,k_{j}}\}\nonumber\\
&+\left([v_{\ell}]_r\right)^2\bigg\}\Bigg).
\end{align}
The second term of right side represents the cross product of different parameters which is small compared to other terms. By neglecting this term, it can be verified that 
\begin{align}
\Ex(\mathcal{G}) & =ML-2ML\mu_r+ML\Ex(\left([\lambda]_r\right)^{2})+L\sigma_{\vartheta}^{2}/2\label{eq:eg4}
\end{align}
 Hence, by setting $\Ex(\mathcal{G}) =[\mathbf{y}]_r^{T}[\mathbf{y}]_r$ in \eqref{eq:eg4}
one can obtain a rough estimate of $M$.

\section{Sparse signal recover based approach}
\subsection{Basis mismatched sparse representation}
The CMUD algorithm is computationally very efficient but cannot detect large number of active users. Furthermore, CMUD requires the information of variances of channel mismatch parameters i.e., $\sigma_h^2,\sigma_e^2,\sigma_{\eta}^2$. On the other hand, MUD being a sparse recovery problem, can be solved by using sparse recovery algorithms \cite{mud8,mud5}. Although the sparse recovery algorithms are relatively computationally demanding, they have been demonstrated the capability of resolving larger number of active users provided that the information of channel mismatch parameters are available. However, if the information of $\sigma_h^2,\sigma_e^2,\sigma_{\eta}^2$ are unavailable then we cannot efficiently apply the standard sparse recovery algorithms \cite{mud8,mud5} to the data model in \eqref{eq:dat2}. To overcome the channel mismatch limitations, we pose the user detection as a basis mismatched sparse signal recovery problem and develop an efficient algorithm to resolve the MUD problem. To understand the concept of basis mismatch, rewrite the data model in \eqref{eq:dat2} as: $$\zy-\vartheta= [ \ \bC-\bQ \ ]\mathring{\zx}$$ where we need to estimate $\mathring{\zx}$. Since $\mathring{\zx}$ is sparse, $\zy-\vartheta$ is a linear combination of few selected columns of $[ \ \bC-\bQ \ ]$. Hence, we say that $\zy-\vartheta$ has a sparse representation on the basis $[ \ \bC-\bQ \ ]$.
The basis mismatch problem arises when we do not have accurate knowledge about a part of the basis. In the present scenario we do not have actual information about $\bQ$. 

\subsection{Solution strategy}\label{sec:tls}
In this work, we adopt a total least-squares (TLS) optimization with sparsity constraint \cite{bmis5} to handle the basis mismatch problem and obtain an estimate of $\mathring{\zx}$. The sparsity constraint TLS algorithm aims to solve the following optimization \cite{bmis5}:
\begin{align}\label{eq:tls1}
\{\zx_T,\bQ_T\}&=\arg \ \min_{\zx,\bQ} \ G_p(\zx,\bQ)\\
\mathrm{where,}\ G_p(\zx,\bQ)&=\|[\zy]_r-(\bC-\bQ)\zx\|_2^2+\|\bQ\|_F^2+\frac{2}{\xi}\|\zx\|_p\nonumber
\end{align}
where $0\le p\le 1$ and the value of $\xi$ depends on noise level. The sparsity constraint in \eqref{eq:tls1} has been imposed by the $\ell_p$ norm ($0\le p\le 1$). The optimization problem is nonconvex and no effective optimization solvers is available that can guarantee the convergence to the global minimum of \eqref{eq:tls1}. To handle the problem, the work in \cite{bmis5} splits the optimization into two parts and develop an iterative algorithm to resolve $\bQ$ and $\zx$ in alternative fashion. In particular, given the iterate $\bQ^{(i)}$ at iteration $i\ge0$, the iterate $\zx^{(i)}$ is obtained by solving the following optimization
\begin{align}\label{eq:opt1}
\zx^{(i)}=\arg \min_{\zx} G_p(\zx,\bQ^{(i)}).
\end{align}  
With $\zx^{(i)}$ available, $\bQ^{(i+1)}$ is found as
\begin{align}\label{eq:opt2}
\bQ^{(i+1)}=\arg \min_{\bQ} G_p(\zx^{(i)},\bQ).
\end{align}
The procedure is repeated at every iteration until a convergence criterion is satisfied. Note that the second optimization has a closed form of solution \cite{bmis5}:
\begin{align}\label{eq:Qup}
\bQ^{(i+1)}=\left(1+\|\zx^{(i)}\|_2^2\right)^{-1}[\bC\zx^{(i)}-[\zy]_r](\zx^{(i)})^\top.
\end{align}
The work in \cite{bmis5} sets $p=1$, which results in a convex optimization problem in \eqref{eq:opt1}.
However, it has been shown in \cite{rls,foc} that some nonconvex relaxations of $\ell_1$ norm i.e., $(0<p<1)$ are more efficient in finding sparse solutions. Furthermore, they take lower number of iterations compared to the $\ell_1$ based optimization. 
Again, in the present case $\mathring{\zx}_i\in\{1,0\}$ which has not been considered in the optimization \eqref{eq:tls1}.
In this section, we first adopt a nonconvex relaxation of $\ell_1$ norm, i.e, $\ell_p$ norm $(0<p<1)$ for the optimization in \eqref{eq:opt1}. Second, we incorporate the constraint  $\mathring{\zx}_i\in\{1,0\}$ with the optimization. Finally, we propose an algorithm that applied an iterative reweighted least-square technique to minimize a sequence like \eqref{eq:tls1}. We shall demonstrate by simulations that the optimization yields better user detection.

The optimization problem in \eqref{eq:opt1} is nonconvex for $0<p<1$. The work in \cite{foc} suggests an iterative algorithm called FOCUUS to solve the optimization. At $j$-th iteration, the algorithm solves a reweighted $\ell_2$ minimization problem:
\begin{align}\label{eq:foc2}
\bar{\zx}^{(j)}&=\arg \min_{\zx} S_p^{(j)}(\zx,\bQ^{(i)})\\
\mathrm{with,} \ \ \ \ \ &\nonumber\\
 \ S_p^{(j)}(\zx,\bQ^{(i)})&=\frac{\xi}{2}\|[\zy]_r-(\bC-\bQ^{(i)})\zx\|_2^2+\frac{1}{2}\zx^{\top}\bW_p^{(j-1)}\zx\nonumber
\end{align}
where $\bW_p^{(j-1)}=\mathrm{diag}\{|\bar{\zx}_1^{(j-1)}|^{p-2},\cdots ,|\bar{\zx}_K^{(j-1)}|^{p-2}\}$. The \eqref{eq:foc2} is a quadratic optimization problem and has closed form of solution. Starting from an initial estimate $\bar{\zx}^{(0)}$, the algorithm update $\bar{\zx}^{(j)}$ in every iteration. The algorithm terminates when $|\bar{\zx}^{(j)}-\bar{\zx}^{(j-1)}|$ becomes small. It has been shown in \cite{foc,fochy} that the cost function $S_p^{(j)}(\zx,\bQ^{(i)})$ is monotonically non-increasing on the sequence $\{\bar{\zx}^{(j)}\}_{j=1}^{\infty}$. Furthermore, the limit of any convergent
subsequence of $\{\bar{\zx}^{(j)}\}_{j=1}^{\infty}$ is a stationary point of the problem in \eqref{eq:opt1}.

To incorporate the constraint $\mathring{\zx}_i\in\{0,1\}$ with every iteration of the FOCUUS algorithm, we consider the following optimization problem:
\begin{align}\label{eq:foc3}
\hat{\zx}^{(j)}&=\arg \min_{\zx}  S_p^{(j)}(\zx,\bQ^{(i)})\\
\label{eq:bin} &\mathrm{ Sub.~ to.} \ \zx_t(\zx_t-1)=0; \ \ t=1,2,\cdots K.
\end{align}
Due to the binary constraint in \eqref{eq:bin}, the primal problem above is not convex. Hence, we propose a Lagrangian Dual Relaxation method to optimize \eqref{eq:foc3}. The Lagrangian associate to \eqref{eq:foc3}-\eqref{eq:bin} is
\begin{align}\label{lag1}
\mathcal{L}(\zx,\gmu)&=\zx^{\top}\bP_p^{(j-1)}\zx -(\xi[\zy]_r^{\top}(\bC-\bQ^{(i)})+\gmu^{\top})\zx\\
\mathrm{where,}& \ \bP_p^{(j-1)}=\nonumber\\
&\left(\bW_p^{(j-1)}+\frac{\xi}{2}(\bC-\bQ^{(i)})^{\top}(\bC-\bQ^{(i)})+\mathrm{diag}\{\gmu\}\right)
\end{align}
and $\gmu\in\mathbb{R}^K$ is the dual variable. The dual function of \eqref{lag1} is $g(\gmu)=\inf_{\zx}\mathcal{L}(\zx,\gmu).$ Since $\mathcal{L}(\zx,\gmu)$ is a convex function of $\zx$, we can find the minimizing $\zx$ from the optimality condition $\frac{\partial \mathcal{L}(\zx,\gmu)}{\partial \zx}=0$, which yields
\begin{align}\label{eq:zx}
\zx = \frac{1}{2}\left(\bP_p^{(j-1)}\right)^{-1}(\xi(\bC-\bQ^{(i)})^{\top}[\zy]_r+\gmu).
\end{align}
Putting the value of $\zx$ in \eqref{lag1}, the dual function is
\begin{align}\label{eq:dual}
g(\gmu)=&\frac{1}{4}\Bigg[\left(\xi[\zy]_r^{\top}(\bC-\bQ^{(i)})-\gmu^{\top}\right)\left(\bP_p^{(j-1)}\right)^{-1}\nonumber\\
& \ \ \ \ \ \ \ \ \ \ \ \ \ \ \ \left(\xi(\bC-\bQ^{(i)})^{\top}[\zy]_r+\gmu\right)\Bigg].
\end{align}
Now we need to find $\gmu$ that maximize \eqref{eq:dual}. Once we find $\gmu$, we can compute $\zx$ using \eqref{eq:zx}. We apply a gradient ascent method to maximize \eqref{eq:dual}. The method requires to compute the gradient of \eqref{eq:dual} with respect to $\gmu$.  
At first note that for any vectors $\zr$ and $\zs$ independent from $\gmu$, we have
\begin{align}\label{eq:dif}
&\frac{\partial [\zr^{\top} \left(\bP_p^{(j-1)}\right)^{-1} \zs]}{\partial \gmu_t}\nonumber\\
&=\mathrm{Tr}\left\{\left(\frac{\partial [\zr^{\top} \left(\bP_p^{(j-1)}\right)^{-1} \zs]}{\partial \left(\bP_p^{(j-1)}\right)}\right)\frac{\partial  \bP_p^{(j-1)}}{\partial \gmu_t}\right\}
\end{align}
where
\begin{align}
\frac{\partial [\zr^{\top} \left(\bP_p^{(j-1)}\right)^{-1} \zs]}{\partial \left(\bP_p^{(j-1)}\right)}=-\left( \left(\bP_p^{(j-1)}\right)^{-1}\zs  \zr^{\top} \left(\bP_p^{(j-1)}\right)^{-1}\right)^{\top}
\end{align}
By applying chain rule and using \eqref{eq:dif}, it can be verified that the first order derivative of \eqref{eq:dual} with respect to $\gmu$ is
\begin{align}\label{eq:grad}
\frac{\partial g(\gmu)}{\partial \gmu}&=\frac{1}{4}\Big[-\mathrm{Diag}\left( \left(\bP_p^{(i-1)}\right)^{-1}\zz \ \zv^{\top} \left(\bP_p^{(i-1)}\right)^{-1}\right)\nonumber\\
& \ \ -\left(\bP_p^{(i-1)}\right)^{-1}\zv+\left(\bP_p^{(i-1)}\right)^{-1}\zz\Big]\\
\mathrm{where}, & \nonumber\\
 \ \zz=(\xi &(\bC-\bQ_*)^{\top}[\zy]_r-\gmu); \ \zv=(\xi(\bC-\bQ_*)^{\top}[\zy]_r+\gmu)\nonumber
\end{align}
and $\mathrm{Diag}(\bA)$ denotes a vector constructed from the diagonal components of the matrix $\bA$. 

By using the results, we are now ready to develop an iterative algorithm to solve the basis mismatched sparse recovery problem. 
Motivated by the objectives of \eqref{eq:tls1} and \eqref{eq:foc2}, the algorithm aims to find the sequence $\{\hat{\zx}^{(i)},\hat{\bQ}^{(i)}\}_{i=1}^{\infty}$ that will satisfy: 
\begin{align}\label{eq:falg}
\tilde{G}_p(\hat{\zx}^{(i)},\hat{\bQ}^{(i)})&\le \tilde{G}_p(\hat{\zx}^{(i-1)},\hat{\bQ}^{(i-1)})\\
\mathrm{where,} \ \tilde{G}_p(\hat{\zx}^{(i)},\hat{\bQ}^{(i)}) =&\nonumber\\
\|[\zy]_r-(\bC-\hat{\bQ}^{(i)})\hat{\zx}^{(i)}\|_2^2&+\|\hat{\bQ}^{(i)}\|_F^2+\frac{1}{\xi}[\hat{\zx}^{(i)}]^{\top}\bW_p^{(i-1)}\hat{\zx}^{(i)}
\end{align}
with the constraint in \eqref{eq:bin}. Note that the main difference of the objective in \eqref{eq:falg} with \eqref{eq:tls1} is that instead of $\ell_1$ norm, we incorporate a reweighted quadratic term in \eqref{eq:falg}.
Similar to \cite{bmis5}, we split \eqref{eq:falg} into two parts. Given the iterate $\hat{\bQ}^{(i-1)}$ the algorithm first minimizes $\tilde{G}_p(\zx,\hat{\bQ}^{(i-1)})$ for $\hat{\zx}^{(i)}$ with constraint \eqref{eq:bin}. Once $\hat{\zx}^{(i)}$ is obtained, the algorithm solves $\hat{\bQ}^{(i)}$ by using \eqref{eq:Qup}. The final $\ell_p$-constrained TLS algorithm is given in Table-\ref{tab:algo2}. 
In Step-2, we compute $\gmu$ by applying a standard gradient ascent algorithm \cite{co}. In step-3, we compute $\hat{\zx}^{(i)}$ using \eqref{eq:zx}. Note that Step-2 ensures that $S_p^{(i)}(\hat{\zx}^{(i)},\hat{\bQ}^{(i-1)})\le S_p^{(i-1)}(\hat{\zx}^{(i-1)},\hat{\bQ}^{(i-1)})$. By using the fact, it can be verified that the algorithm satisfies $\tilde{G}_p(\hat{\zx}^{(i)},\hat{\bQ}^{(i)})\le \tilde{G}_p(\hat{\zx}^{(i-1)},\hat{\bQ}^{(i-1)})$ at every iteration.


\begin{table}[!t]
\renewcommand{\arraystretch}{1.3}
\centering \caption{$\ell_p$-TLS Algorithm}\label{tab:algo2}
\begin{tabular}{l}
 \hline
{\bf Initialization}\\
\ \ \ 1. Set   $\hat{\bQ}^{(0)}={\bf 0}, \hat{\zx}^{(0)}=\bC^{\top}(\bC\bC^{\top})^{-1}\zy$ and $ \nu\in[0,1]$.\\
{\bf For} $i=1,2,\cdots $ do\\
\ \ \ 2. By using the gradient of $g(\gmu)$ in \eqref{eq:grad}, apply a gradient ascent\\
\ \ \ \ \ \ method to find $\gmu$ that maximize $g(\gmu)$ and satisfies \\
\ \ \ \ \ \ $S_p^{(i)}(\hat{\zx}^{(i)},\hat{\bQ}^{(i-1)})\le S_p^{(i-1)}(\hat{\zx}^{(i-1)},\hat{\bQ}^{(i-1)})$.\\
\ \ \ 3. Compute $\hat{\zx}^{(i)}$ by using \eqref{eq:zx}.\\
\ \ \ 4. Update $\hat{\bQ}^{(i)}$ by using \eqref{eq:Qup}.\\
\ \ \ 5. If $|\hat{\zx}^{(i)}-\hat{\zx}^{(i-1)}|<\nu$ then terminate.\\
{\bf End for} \\
{\bf Output:} Project $\hat{\zx}^{(i)}$ onto $\{0,1\}$.\\
\hline
\end{tabular}
\end{table}

\subsection{Performance analysis of sparse recovery algorithms}
The optimization in \eqref{eq:tls1} is nonconvex, hence it is difficult to analyse the user detection performance of the sparse TLS algorithm. However, if the information of variances of channel mismatch parameters i.e., $\sigma_h^2,\sigma_e^2,\sigma_{\eta}^2$ are available then we can represent the data model in \eqref{eq:dat2} as
\begin{align}
\zy=\bC\mathring{\zx}+\zw
\end{align}
where $\zw=-\zu+\vartheta$. By using the proof of Lemma-\ref{lem:nz} it can be shown that $[\zw]_r$ has Gaussian distribution with covariance matrix $\sigma_w^2\bI$ where $\sigma_w^2=M\sigma_r^2+\sigma_{\vartheta}^2$. Then we can apply Lasso to recover $\mathring{\zx}$:
\begin{align}
\zx_*=\arg \min_{\zx} \frac{1}{2}\|[\zy]_r-\bC\zx\|_2^2+\frac{1}{\xi}\|\zx\|_1.
\end{align}
Define the set $\mathbb{L}=\{i\in\{1,2,\cdots K\}:|\zx_{*i}|>0\}$. The following lemma obtained from \cite{mud5} describes the user detection performance of Lasso.
\begin{Lem}\label{lem2}Suppose that $M<L$ and the regularization parameter $\xi$ satisfies $\xi<\frac{c_1}{\sqrt{2\sigma_w^2L\log(K)}}$. Then the Lasso will produce a unique solution $\zx_*$ satisfying $\mathbb{L}\subseteq\mathcal{S}$ with probability greater than $1-4\exp(-c_2L^2/\xi^2)$ where $c_1,c_2$ are some constants whose values depend on the matrix $\bC$.
\end{Lem}

\section{Simulation \& Results}

To evaluate the performance of the proposed algorithms, we develop a Monte
Carlo simulation model using MATLAB. The simulation scenarios are on a WiMAX network based on the OFDMA/TDD profile.
The key simulation parameters are outlined in Table I. Total $L=144$ subcarriers are reserved for the BR purpose. 
The modulation pulse is a root-raised-cosine function with a roll-off $0.22$ and duration $10T_s$. The BR terminals follow a mixed channel model specified by ITU IMT-2000 standards: Ped-A, Ped-B, and Veh-A. The BR terminals select the channel models with equal probability. 
The mobile speed varies in the interval 
$[0,5]$ m/s for Ped-A, Ped-B channels, and $[5,20]$ m/s for Veh-A. We assume that the variances of channel mismatch noise $e$ and $\eta$ are equal i.e., $\sigma_e^2=\sigma_{\eta}^2$. We follow the direction of \cite{5683407} and set
$\alpha=\textrm{e}^{\mathbf{i}\theta}$ where $\theta=5^{0}$. Four different methods are applied for code detection: i) CMUD with Decoder-I (Section-\ref{sec:dec1}) will be denoted by D-I.  ii) CMUD with Decoder-II (Section-\ref{sec:dec2}) will be denoted by D-II. 
iii) Lasso as recommended in \cite{mud5}, and iv) $\ell_p$ (with $p=0.5$) constrained TLS (Section-\ref{sec:tls}) will be denoted by TLS. 

\begin{table}
\caption{Simulation Parameters}

\noindent \centering{}%
\begin{tabular}{ccc}
\hline 
{\footnotesize{Parameters}} & {\footnotesize{Notation}} & {\footnotesize{Values}}\tabularnewline
\hline 
\hline 
{\footnotesize{Base Frequency }} & {\footnotesize{$f_{sys}$}} & {\footnotesize{2.3 GHz}}\tabularnewline
{\footnotesize{Channel Bandwidth }} & {\footnotesize{$f$}} & {\footnotesize{5 MHz}}\tabularnewline
{\footnotesize{Sampling Frequency}} & {\footnotesize{$f_{n}$}} & {\footnotesize{5.6 MHz}}\tabularnewline
{\footnotesize{FFT Size }} & {\footnotesize{$N$}} & {\footnotesize{512}}\tabularnewline
{\footnotesize{Subcarrier Spacing }} & {\footnotesize{$\triangle f$}} & {\footnotesize{10.94 KHz}}\tabularnewline
{\footnotesize{OFDM symbol time }} & {\footnotesize{$T_{s}$}} & {\footnotesize{91.4 $\mu s$}}\tabularnewline
{\footnotesize{Cyclic Prefix (CP) Time }} & {\footnotesize{$T_{cp}$}} & {\footnotesize{$T_{s}/8$}}\tabularnewline
{\footnotesize{No. of CP Samples}} & {\footnotesize{$N_{cp}$}} & {\footnotesize{64}}\tabularnewline
{\footnotesize{Frame Duration }} & {\footnotesize{$T_{f}$}} & {\footnotesize{5 ms}}\tabularnewline
\hline 
\end{tabular}
\label{tab:env}
\end{table}

\begin{figure}[!h]
\centering
\includegraphics[width=8.5cm]{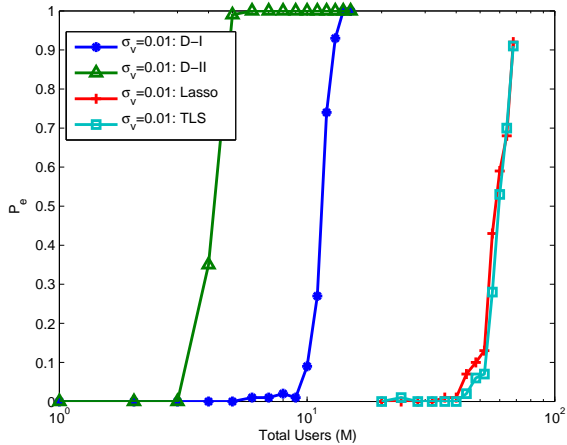}
\caption{Performance of different algorithms with low noise variances. We set $\sigma_e=0.01,\sigma_{\vartheta}=0.01$ and $K=256$. } 
\label{fig1}
\end{figure}
At first we evaluate the code detection performance by different algorithms at low noisy environment in Figure-\ref{fig1}. This simulation will illustrate the capability of detecting largest number of active users by different algorithms. As can be seen, the D-II shows worst user detection performance whereas D-I performs much better than D-II. For example, with $P_e\le0.1$, the D-II can detect only $3$ users whereas D-I is capable of detecting upto $10$ users. As being claimed, Lasso and TLS performs much better than correlation based approach. The TLS can detect $52$ users with $P_e=0.1$ which is almost $5$ times larger than D-I. Finally the figure also reveal the fact that, at low noisy environment, both Lasso and TLS performs similarly.

\begin{figure*}[]
\centerline{\subfloat[]{\includegraphics[width=8.5cm]{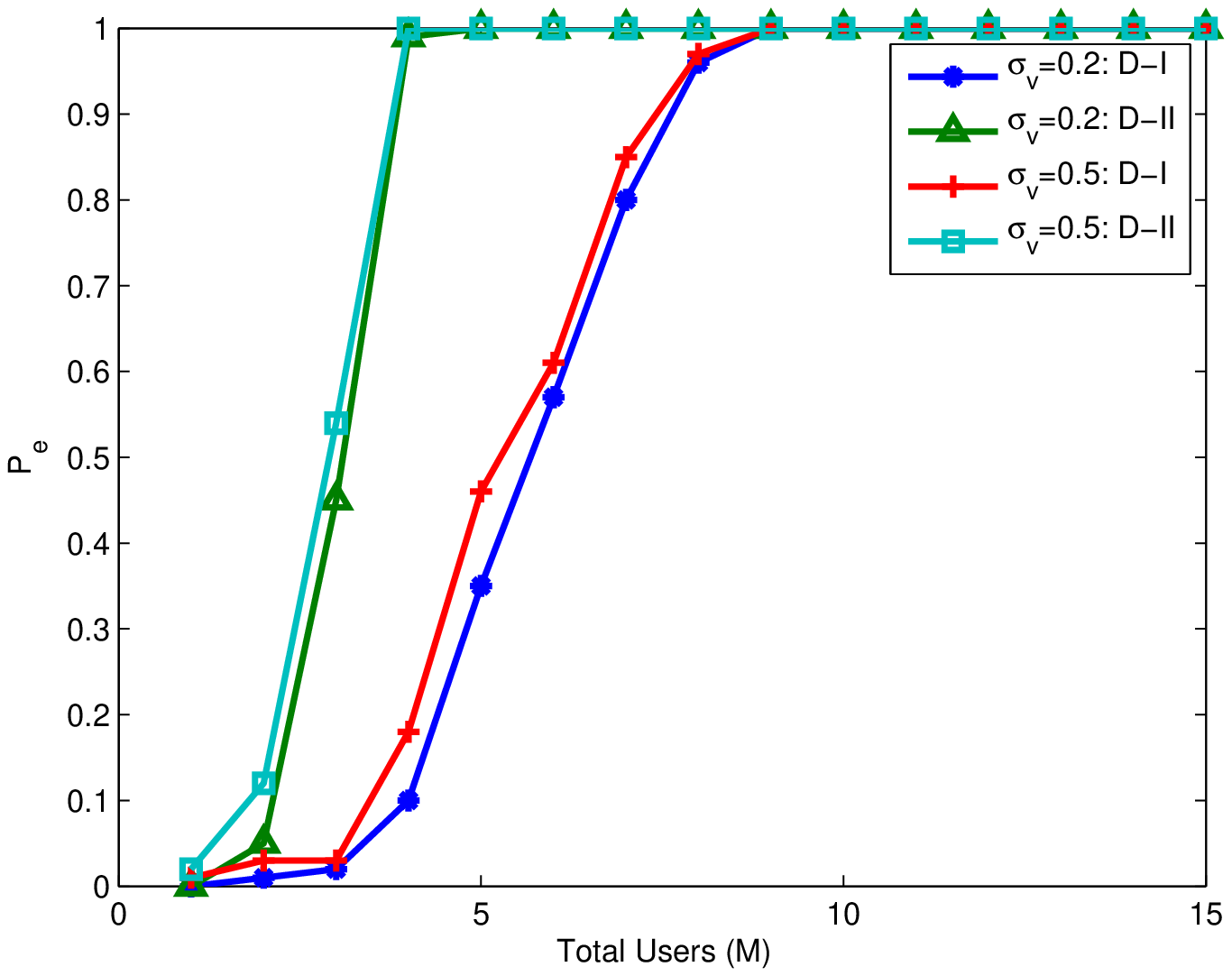}
\label{fig21}}
 \hfil \subfloat[]{\includegraphics[width=8.5cm]{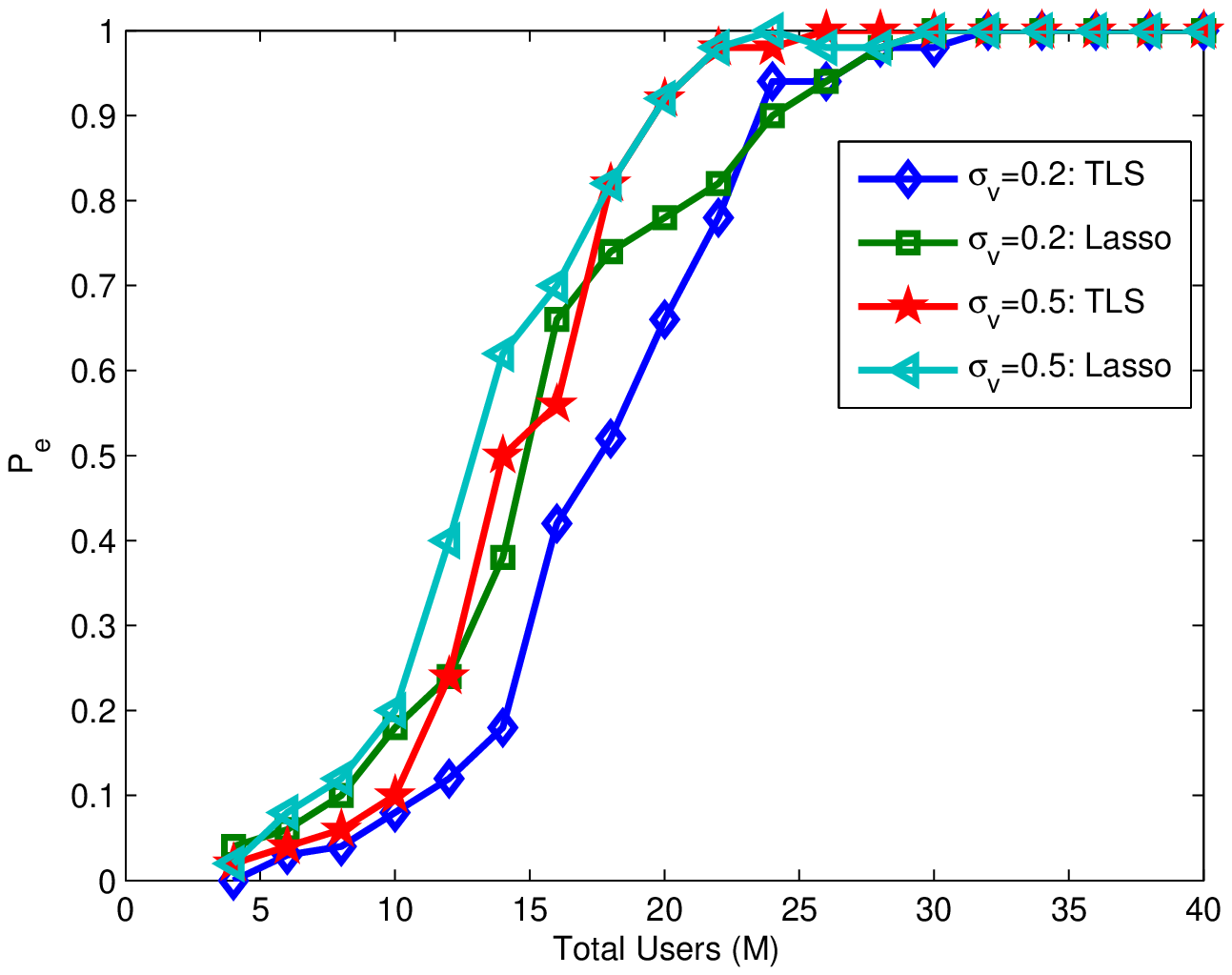}
\label{fig22}}} \caption{ Code detetction performance at differnt measuremnt noise levels $\sigma_{\vartheta}$. We set $\sigma_e=0.15$,  $\sigma_{\eta}=0.15$ and $K=256$.  } \label{fig2}
\end{figure*}

We now investigate performances of different algorithms at high noisy environment. Figure-\ref{fig2} illustrates performances of different algorithms with measurement noise variance $\sigma_{\vartheta}^2$. Figure-\ref{fig2}(a) compares performance of CMUD decoders. The performance of a particular algorithm changes slowly with $\sigma_{\vartheta}$. For example, with $M=4$ users the $P_e$ of D-I are $0.1$ and $0.18$ for $\sigma_{\vartheta}=0.2$ and $0.5$ respectively. Again $P_e$ are $0.57$ and $0.61$ for $\sigma_{\vartheta}=0.2$ and $0.5$ respectively with $M=6$ users. As before, the performance of D-II is worse compared to other algorithms. With $P_e\le0.12$, the D-II can detect only $2$ users.
Furthermore, comparing Figure-\ref{fig2}(a) with Figure-\ref{fig1} we see that the user detection performances of all algorithms drop with increasing noise variances. 
Figure-\ref{fig2}(b) compares performances between Lasso and TLS. As can be seen the code detection performance of both algorithms decreases with increasing $\sigma_{\vartheta}$. For TLS with $M=12$, the $P_e$ are $0.12$ and $0.24$ for $\sigma_{\vartheta}=0.2$ and $0.5$ respectively. The code detection performance gap increases with increasing the number of total active users. For example, with $M=20$ the $P_e$ are $0.66$ and $0.92$ for $\sigma_{\vartheta}=0.2$ and $0.5$ respectively. The performance of Lasso also changes with $\sigma_{\vartheta}$. With $M=8$ the $P_e$ are $0.1$ and $0.12$ for $\sigma_{\vartheta}=0.2$ and $0.5$ respectively.  Note that while  Figure-\ref{fig1} shows that both Lasso and TLS performs similarly al low noisy environment, their performance do not remain similar at high noisy environment in Figure-\ref{fig2}(b). 
 
\begin{figure*}[]
\centerline{\subfloat[]{\includegraphics[width=8.5cm]{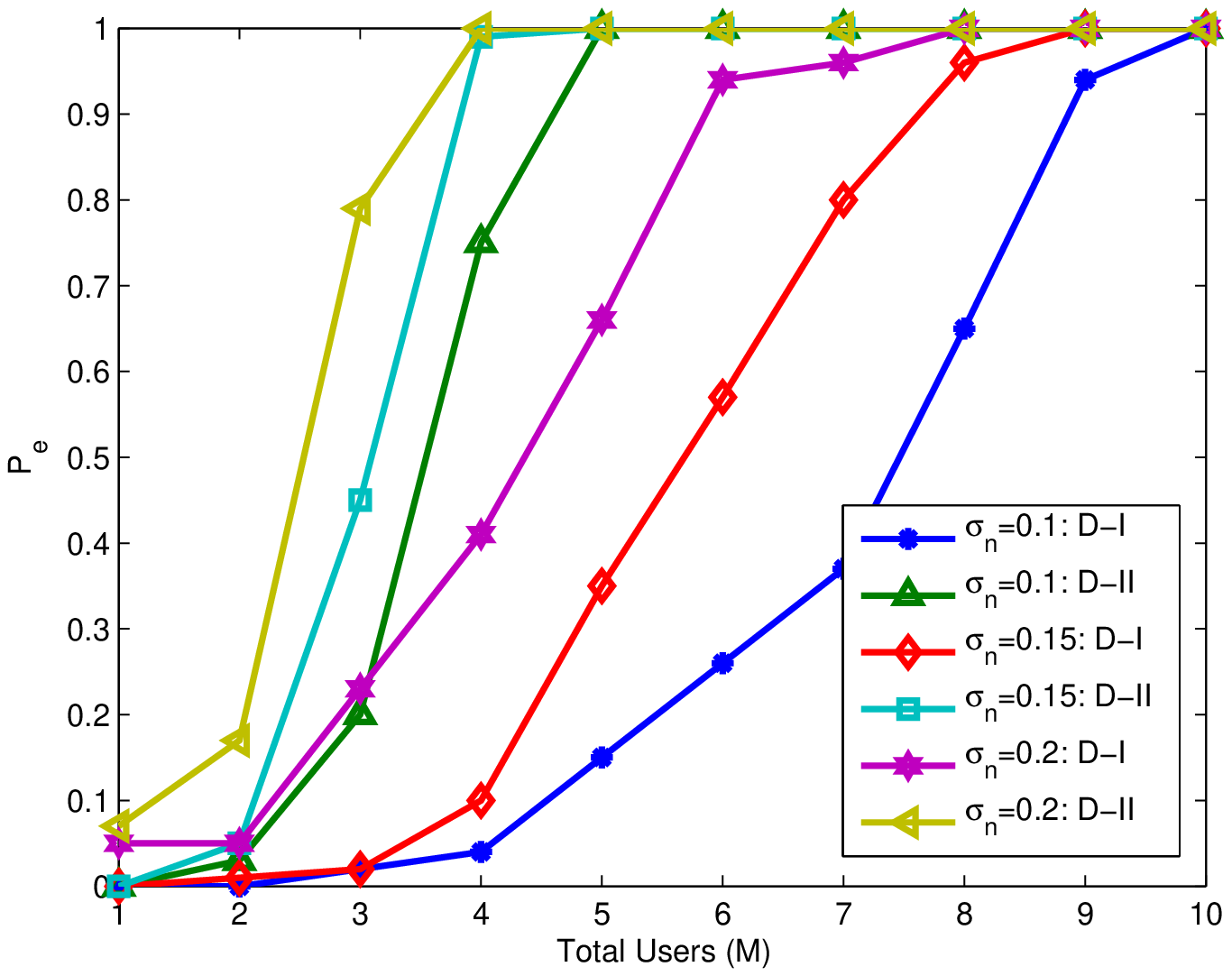}
\label{fig31}}
 \hfil \subfloat[]{\includegraphics[width=8.5cm]{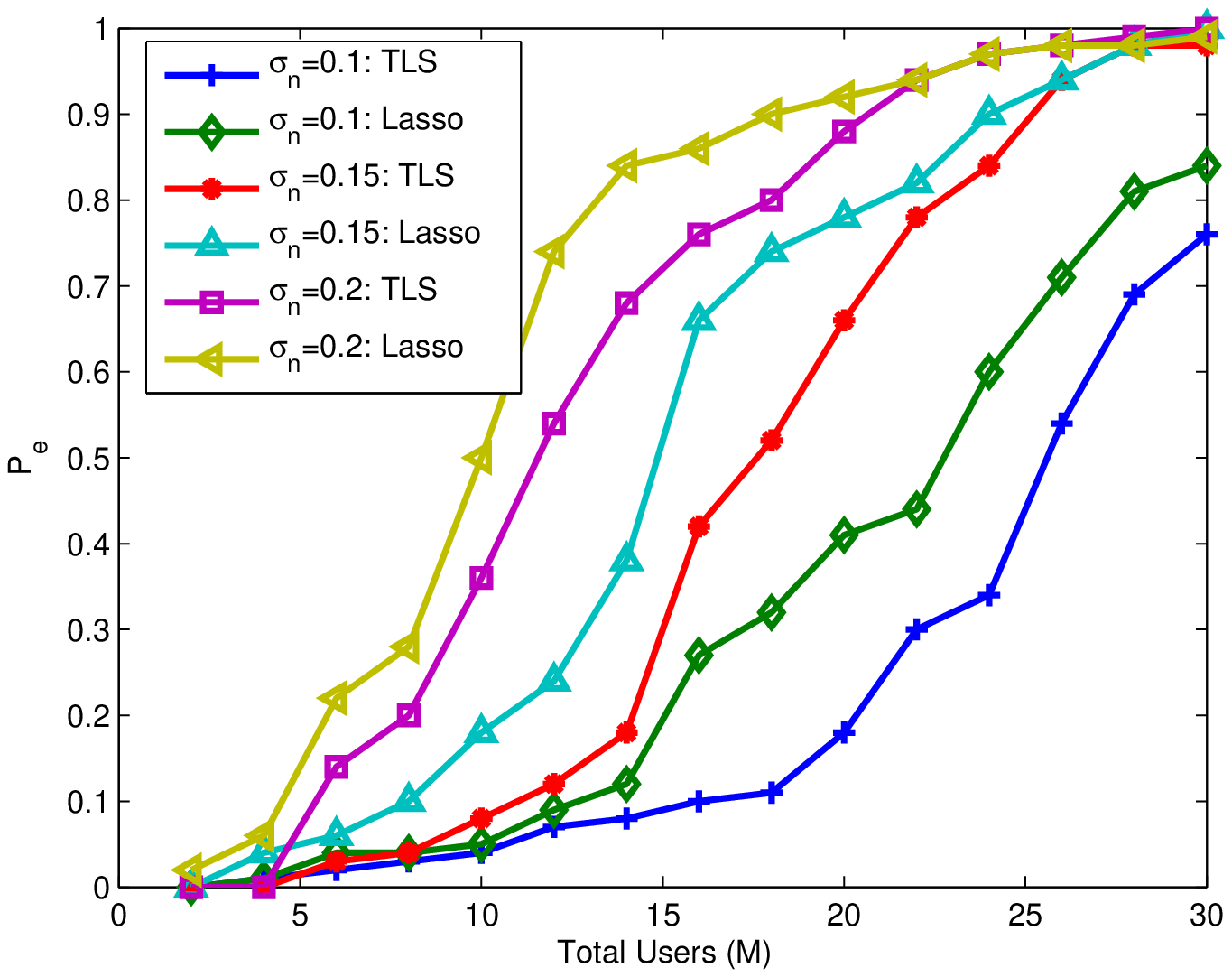}
\label{fig32}}} \caption{ Code detetction performance at differnt channel mismatch noise levels $\sigma_{e}$ and  $\sigma_{\eta}$ where $\sigma_{e}=\sigma_{\eta}$. We set $\sigma_{\vartheta}=0.2$ and $K=256$.  } \label{fig3}
\end{figure*}
Figure-\ref{fig3} shows the effect of channel mismatch on the performance of code detection by different algorithms. Figure-\ref{fig3}(a) shows that the performance of CMUD decreases rapidly with increasing $\sigma_{\eta}$. For D-II with $M=3$, the value of $P_e$ are $0.2$   $0.45$ and $0.79$ for $\sigma_{\eta}=0.1,0.15$ and $0.2$ respectively. The performance of D-I is relatively better than D-II. For example with $M=3$ the value of $P_e$ are $0.02$   $0.025$ and $0.23$ for $\sigma_{\eta}=0.1,0.15$ and $0.2$ respectively. The performance of D-I degrades rapidly at $\sigma_{\eta}=0.2$. In contrast, the TLS performs much better than CMUD decoders. In Figure-\ref{fig3}(c) we see that with $M=12$ the value of $P_e=0.07,0.12$ and $0.54$ receptively for  $\sigma_{\eta}=0.1,0.15$ and $0.2$. In a similar setting (i.e., with $M=12$), the value of $P_e=1$ for D-I (see Figure-\ref{fig3}(a)). Figure-\ref{fig3}(b) also exhibits that TLS performs better than Lasso. With $M=12$ the value of $P_e$ for Lasso are $0.09,0.24$ and $0.74$ receptively for  $\sigma_{\eta}=0.1,0.15$ and $0.2$.

\begin{figure*}[]
\centerline{\subfloat[]{\includegraphics[width=8.5cm]{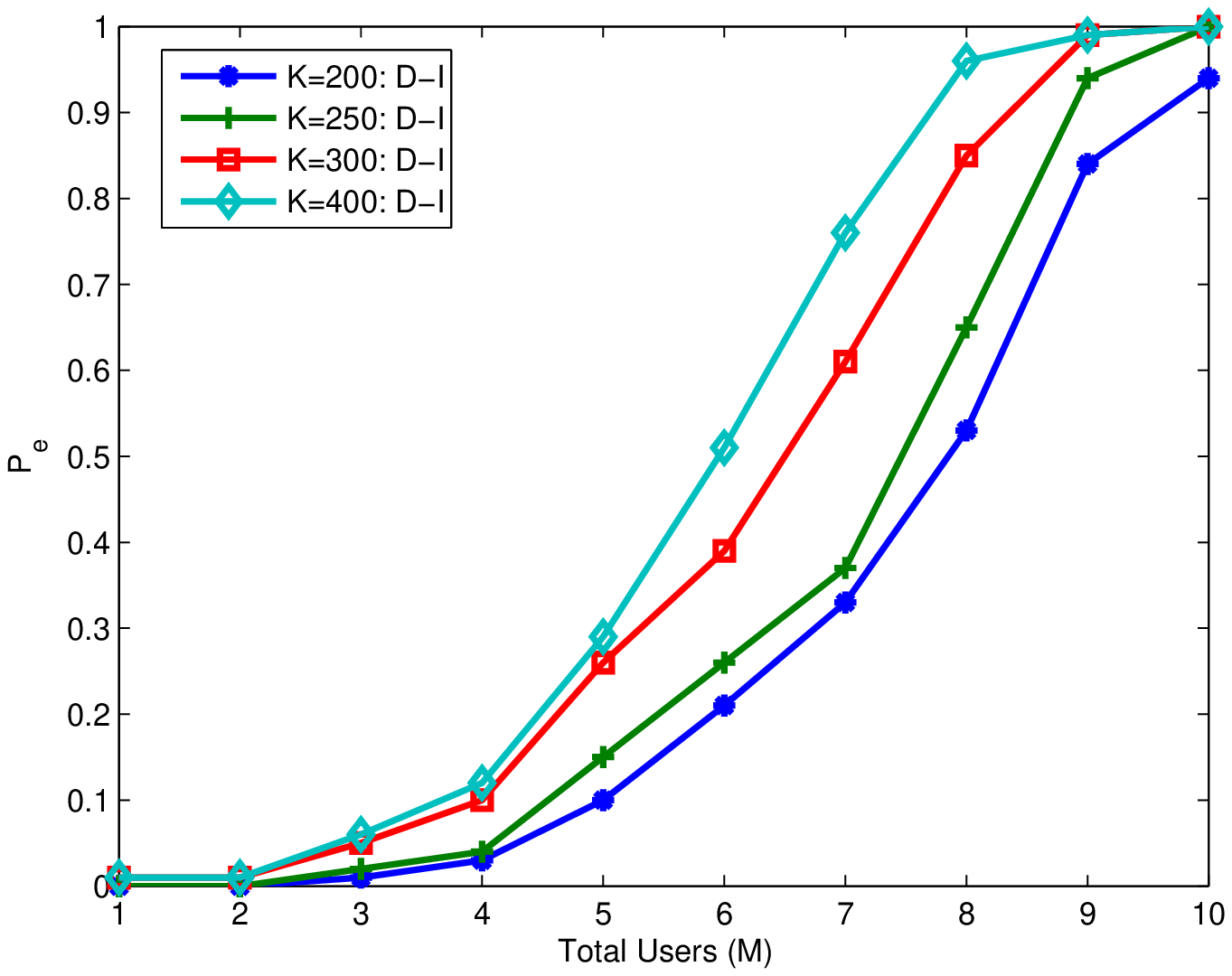}
\label{fig31}}
 \hfil \subfloat[]{\includegraphics[width=8.5cm]{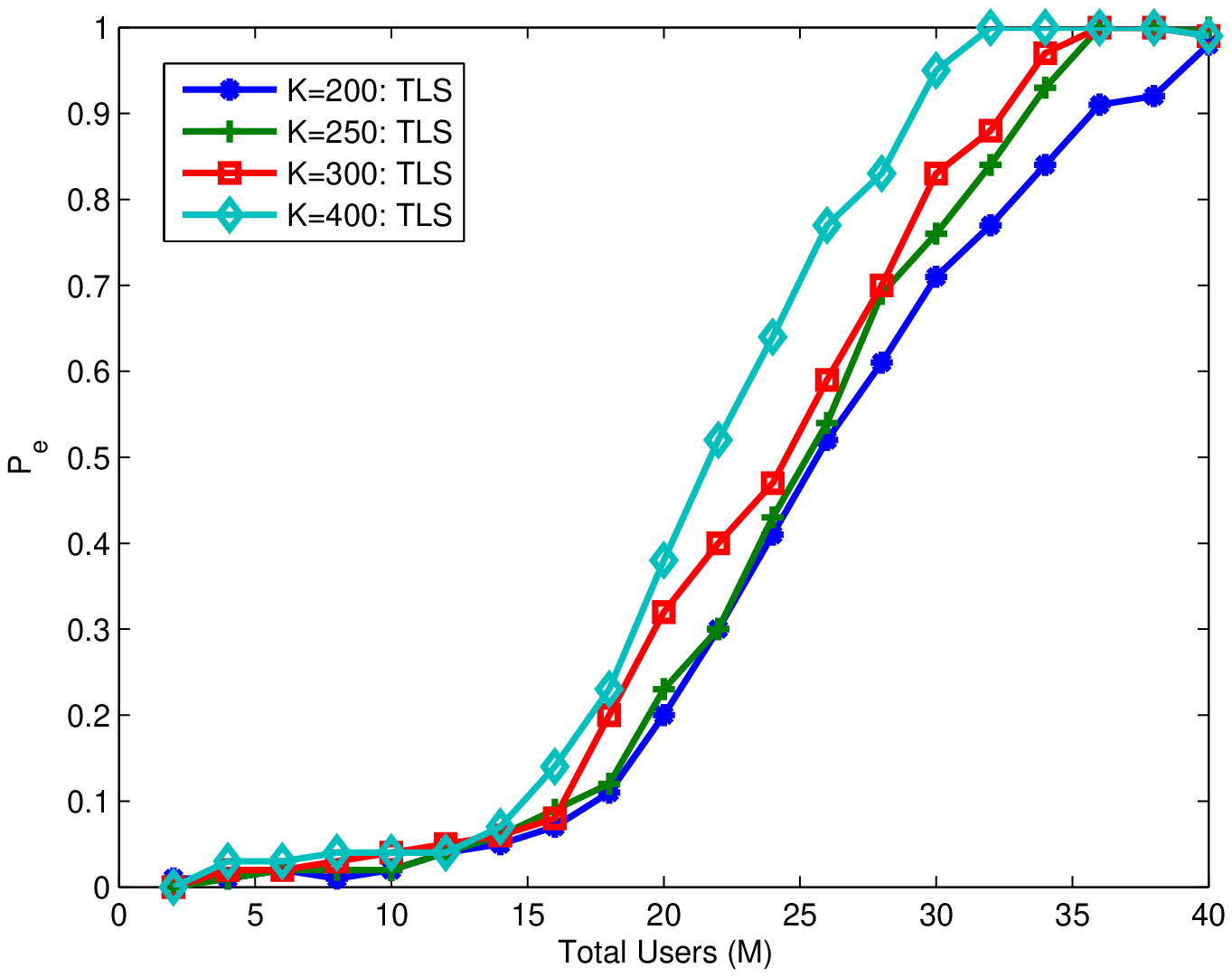}
\label{fig32}}} \caption{ Code detetction performance for differnt values of $K$. (a) CMUD with D-I (b) TLS. We set $\sigma_e=0.1,\sigma_{\eta}=0.1,\sigma_{\vartheta}=0.2$. } \label{fig4}
\end{figure*}

In Section-\ref{sec:sp}, we have assumed that any BR code will be used by at most one BR subscriber and hence $\mathring{\zx}_i\in\{0,1\}$. This assumption will be valid with high probability when the number of active BR users is much smaller than the number of available BR codes i.e., $M\ll K$. According to the simulation setting in Figure-\ref{fig3}(b) the TLS can detect almost $M=20$ users with high probability  when $\sigma_e=0.1,\sigma_{\eta}=0.1,\sigma_{\vartheta}=0.2$ and $K=256$. As the value of $M$ becomes larger we should increase the value of $K$ to remain consistent with the above assumption. To this aim, we investigate the user detection performance of D-I and TLS as a function of $K$ in Figure-\ref{fig4}. As can be seen, the TLS can detect $18$ users with $P_e=0.11,0.12,0.2$ and $0.23$ for $K=200,250,300$ and $400$ respectively.

\section{Conclusion}

Multiuser detection in random access channels plays an important role in OFDMA wireless network to support services with bursty traffic. However, the existing algorithms suffer from low efficiency when the base station does not have accurate knowledge about the channel frequency response of active users. In this work, two different solutions have been proposed to alleviate the difficulty. The first solution applied a very simple correlation based algorithm. The theoretical analysis of the algorithm provides a direction to achieve optimum performance from the algorithm. The second solution applied a variant of sparse recovery algorithm which exhibits better user detection performance, however, with additional computational complexity.

\appendices{}

\section{Mathematical Variance of $\lambda$}\label{app1}

Denote the covariance matrix of $[[\lambda]_{r},\ [\lambda]_{i}]^{\top}$ as
\begin{align}
\mathcal{V}([\lambda]_{r},[\lambda]_{i})=\left[\begin{array}{cc}
\sigma_{r}^{2} & \sigma_{ri}\\
\sigma_{ri} & \sigma_{i}^{2}
\end{array}\right],\label{eq:varlam}
\end{align}
where $\sigma_{r}^{2}=\Ex([\lambda]_{r}^{2})-(\Ex([\lambda]_{r}))^{2}$,
$\sigma_{ri}=\Ex([\lambda]_{r}[\lambda]_{i})-\Ex([\lambda]_{r})\Ex([\lambda]_{i})$.
Then $\Ex([\lambda]_{r})$ can be calculated as 
\begin{align}
\Ex([\lambda]_{r})=\int_{-\infty}^{\infty}\int_{-\infty}^{\infty}[\lambda]_{r}f([\lambda]_{r},[\lambda]_{i})d[\lambda]_{r}d[\lambda]_{i}\label{eq:exp}
\end{align}
We evaluate the integral in polar coordinate. Let us define
\begin{align}
\gamma & =\sqrt{\frac{(1-\rho_{r}^{2}-\rho_{i}^{2})\sigma_{u}^{2}}{\sigma_{v}^{2}}};\nonumber \\
\label{eq:lr}[\lambda]_{i} & =t\sin\theta-\frac{\rho_{i}\sigma_{u}}{\sigma_{v}},\ [\lambda]_{r}=t\cos\theta+\frac{\rho_{r}\sigma_{u}}{\sigma_{v}}; \\
\hat{\alpha} & =\frac{\rho_{r}\sigma_{u}}{\sigma_{v}},\beta=\frac{-\rho_{i}\sigma_{u}}{\sigma_{v}},\ \Gamma=\frac{(1-|\rho|^{2})\sigma_{u}^{2}}{\pi\sigma_{v}^{2}}.\label{eq:lim}
\end{align}
Then by combining \eqref{eq:exp}-\eqref{eq:lim} along with the pdf in \eqref{eq:pdf} we get 
\begin{align}
\Ex([\lambda]_{r}) & =\Gamma\int_{0}^{2\pi}\int_{0}^{\infty}\left(\frac{t^{2}\cos\theta}{(t^{2}+\gamma^{2})^{2}}+\frac{t\hat{\alpha}}{(t^{2}+\gamma^{2})^{2}}\right)dtd\theta\nonumber \\
 & =\Gamma\int_{0}^{2\pi}\left[-\left(\frac{\pi}{4\gamma}\right)\cos\theta+\frac{\hat{\alpha}}{2\gamma^{2}}\right]d\theta\nonumber \\
 & =\frac{\Gamma \pi \hat{\alpha}}{\gamma^2}=\hat{\alpha}.
\end{align}
Next we compute $\Ex([\lambda]_{r}^{2})$, 
\begin{align}
\Ex([\lambda]_{r}^{2})=\int_{-\infty}^{\infty}\int_{-\infty}^{\infty}[\lambda]_{t}^{2}f([\lambda]_{r},[\lambda]_{i})d[\lambda]_{r}d[\lambda]_{i}.\label{eq:exp3}
\end{align}
Using the similar conversion in \eqref{eq:lr}-\eqref{eq:lim}, we have 
\begin{align}
\Ex([\lambda]_{r}^{2}) \ \ \ \ & \nonumber\\
= \Gamma\int_{0}^{2\pi}\int_{0}^{\infty}&\left(\frac{t^{3}\cos^{2}\theta}{(t^{2}+\gamma^{2})^{2}}+\frac{2\hat{\alpha}t^{2}\cos\theta}{(t^{2}+\gamma^{2})^{2}}+\frac{t\hat{\alpha}^{2}}{(t^{2}+\gamma^{2})^{2}}\right)dt d\theta\nonumber \\
  =\Gamma\int_{0}^{2\pi}\int_{0}^{\infty}&\left(\frac{t^{3}\cos^{2}\theta}{(t^{2}+\gamma^{2})^{2}}\right)dtd\theta+\frac{\Gamma\pi\hat{\alpha}^{2}}{\gamma^{2}}.\label{eq:lamr}
\end{align}
There is no closed form solution of the first integral term in \eqref{eq:lamr}. In fact,
the solution become unbounded when the upper limiting value of $t\rightarrow\infty$. However,
the solution can be bounded by some limiting value of $t<\infty$. In practice,
$\sigma_{h}$ is bounded and $\sigma_{h}>\sigma_{e},\sigma_{\eta}$,
hence there is very low probability of $([\lambda]_{r},[\lambda]_{i})$
taking very large magnitude. To obtain a reliable bound for $t$, we first find a threshold $\mathcal{T}$ for $([\lambda]_{r},[\lambda]_{i})$ such that both $|[\lambda]_{r}|$ and $|[\lambda]_{i}|$ will remain below $\mathcal{T}$ with high probability. Once we obtain a reliable $\mathcal{T}$, we can compute a bound on $t$ by using \eqref{eq:lr}.
To approximate the value of $\mathcal{T}$, we utilize a concept called percentile level \cite{Papoulis_91}. 
The $\varrho$ percentile level denoted by $\mathcal{T}_{\varrho}$ is defined as:
\begin{align}
\int_{-\mathcal{T}_{\varrho}}^{\mathcal{T}_{\varrho}}\int_{-\mathcal{T}_{\varrho}}^{\mathcal{T}_{\varrho}}f([\lambda]_{r},[\lambda]_{i})d[\lambda]_{r}d[\lambda]_{i}\ge \varrho\label{eq:expnew}
\end{align}
The choice of ${\varrho}$ is dictated by two conflicting requirements: If $\varrho$ is close to $1$, the estimate of threshold $\mathcal{T}_{\varrho}$ is reliable but the sample space of $([\lambda]_{r},[\lambda]_{i})$ is large; if $\varrho$ is small, the sample space is reduced but the estimate is less reliable. In our experiments, we set $\varrho=0.95$.

\section{Proof of Lemma-\ref{lem0}}\label{app2}
Our proof is closely based on the works \cite[Thm. 6]{omp1} and \cite[Thm. 4]{mud6}. Let us define the event
\begin{align}\label{eq:event}
\Sigma=\{\max_{1\le \ell\le K}|[\bD_{\ell}^{\top}(-\zu+\vartheta)]_r|<\tau\}.
\end{align}
We shall demonstrate that $\Sigma$ occurs with high probability and that under \eqref{eq:d} whenever $\Sigma$ occurs, the active codes will be detected correctly.
At first we provide few lemmas which will be used to prove the result.

\begin{Lem} \label{lem:nz}Suppose that $\vartheta$ is a zero mean complex Gaussian random vector with covariance matrix $\sigma_{\vartheta}^2\bI$ and 
$\zu$ is a random vector as defined in \eqref{eq:dat2}. If $(\pi(1+\nu)\log K)^{-1/2}K^{-(1+\nu)}\le 1$ for some $\nu>0$, then the event $\Sigma$ in \eqref{eq:event} occurs with probability at least $1-(\pi(1+\nu)\log K)^{-1/2}K^{-\nu}.$
\end{Lem}
{\bf Proof:}
Note that 
$
[\bD_{\ell}^{\top}\zu]_r$
 is a linear combination of random variables $\{[\lambda_{j,k}]_r\}$ and by using the properties of $\{[\lambda_{j,k}]_r\}$ discussed in Section-\ref{sec:lam}, each variable $[\lambda_{j,k}]_r$ is independent from other with mean $\mu_r$ and variance $\sigma_r^2$. By central limit theorem and using \eqref{eq:mlam}-\eqref{eq:vlam} we see that $[\bD_{\ell}^{\top}\zu]_r$ has a Gaussian distribution with mean $\epsilon\mu_r\sum_{j=1}^K{\bD_{\ell}^{\top}\bC_{j}}$ and variance $M \sigma_r^2\|\bD_{\ell}\|_2^2.$ Furthermore, since $\vartheta$ has zero mean i.i.d. Gaussian distribution and entries of $\vartheta$ are independent from $\zu$, we get that  $[\bD_{\ell}^{\top}(-\zu+\vartheta)]_r$ has Gaussian distribution with mean $-\epsilon\mu_r\sum_{j=1}^K{\bD_{\ell}^{\top}\bC_{j}}$ and variance $M \sigma_r^2\|\bD_{\ell}\|_2^2+\|\bD_{\ell}\|_2^2\sigma_{\vartheta}^2/2.$

The random variables $\{[\bD_{\ell}^{\top}(-\zu+\vartheta)]_r\}_{\ell=1}^K$ are jointly Gaussian. Hence
\begin{align}\label{eq:l11}
Pr(\Sigma)&=Pr(\max_{1\le \ell\le K}|[\bD_{\ell}^{\top}(-\zu+\vartheta)]_r|<\tau)\nonumber\\
&\ge \prod_{\ell=1}^K Pr(|[\bD_{\ell}^{\top}(-\zu+\vartheta)]_r|<\tau)
\end{align}
Denote $\mu_{\ell}=-\epsilon\mu_r\sum_{j=1}^K{\bD_{\ell}^{\top}\bC_{j}}$ and $\sigma_{\ell}^2=M \sigma_r^2\|\bD_{\ell}\|_2^2+\|\bD_{\ell}\|_2^2\sigma_{\vartheta}^2/2.$ By using Gaussian tail bound, we have
\begin{align}\label{eq:l12}
Pr(|[\bD_{\ell}^{\top}(-\zu+\vartheta)]_r|<\tau)\ge 1-\sqrt{\frac{2}{\pi}}\frac{\sigma_{\ell}}{\tau-\mu_{\ell}}e^{\frac{-(\tau-\mu_{\ell})^2}{2\sigma_{\ell}^2}}
\end{align}
By using \eqref{eq:d1}-\eqref{eq:d3} we have $\frac{\sigma_{\ell}}{\tau-\mu_{\ell}}\le \frac{\gamma\sqrt{M\sigma_r^2+\sigma_{\vartheta}^2/2}}{\tau-\epsilon\mu_r\alpha}$. By using the definition of $\tau$ in \eqref{eq:tau},
we see that for any $\ell\in\{1,2,\cdots K\}$:
\begin{align}\label{eq:l13}
\sqrt{\frac{2}{\pi}}\frac{\sigma_{\ell}}{\tau-\mu_{\ell}}e^{\frac{-(\tau-\mu_{\ell})^2}{2\sigma_{\ell}^2}}\le (\pi(1+\nu)\log K)^{-1/2}K^{-(1+\nu)}.
\end{align}
Now combining \eqref{eq:l11}, \eqref{eq:l12}, \eqref{eq:l13} we obtain that
\begin{align}
Pr(\Sigma)\ge \left(1-(\pi(1+\nu)\log K)^{-1/2}K^{-(1+\nu)}\right)^K.
\end{align}
Applying the inequality $(1-x)^K\ge 1-Kx$, valid for $K\ge 1$ and $x\le 1$, we have
\begin{align}
Pr(\Sigma)\ge 1-(\pi(1+\nu)\log K)^{-1/2}K^{-\nu}.
\end{align}
\feop

\begin{Lem}\label{lem2}
Let $\zy=\sum_{\ell\in\mathcal{S}}\bC_{\ell}-\zu+\vartheta$ where the random vector $\vartheta$ has zero mean Gaussian distribution with covariance $\sigma_{\vartheta}^2\bI$ and mean and covariance of the random vector $\zu$ is defined as in \eqref{eq:mlam} and \eqref{eq:vlam} respectively.  Let $\bD\in\mathbb{R}^{L\times K}$ be a decoder matrix such that $\bD_{\ell}^{\top}\bC_{\ell}=1;\forall \ell$. Then under event $\Sigma$, we have
\begin{align}
\max_{j\not\in\mathcal{S}}\{|[\bD_{j}\zy]_r|\}&\le M\beta+\tau\\
\max_{j\in\mathcal{S}}\{|[\bD_{j}\zy]_r|\}&\ge 1- (M-1)\beta-\tau.
\end{align}
\end{Lem}
{\bf Proof:}
At first we see that under the event $\Sigma$,
\begin{align}
\max_{j\not\in\mathcal{S}}\{|[\bD_{j}^{\top}\zy]_r|\}&=\max_{j\not\in\mathcal{S}}\left\{\left|\sum_{\ell\in\mathcal{S}}\bD_{j}^{\top}\bC_{\ell}+\bD_{j}^{\top}([-\zu+\vartheta]_r)\right|\right\}\nonumber\\
\le \max_{j\not\in\mathcal{S}}&\sum_{\ell\in\mathcal{S}}|\bD_{j}^{\top}\bC_{\ell}|+\max_{j\not\in\mathcal{S}}|\bD_{j}^{\top}([-\zu+\vartheta]_r)|\nonumber\\
\label{eq:m1}\le M\beta&+\tau. 
\end{align}
Similarly, under the event $\Sigma$
\begin{align}
\max_{j\in\mathcal{S}}\{|[\bD_{j}^{\top}\zy]_r|\}&\nonumber\\
=\max_{j\in\mathcal{S}}&\left\{\left| 1+\sum_{\ell\in\mathcal{S}\setminus j}\bD_{j}^{\top}\bC_{\ell}+\bD_{j}^{\top}([-\zu+\vartheta]_r)\right|\right\}\nonumber\\
\ge 1- \max_{j\in\mathcal{S}}&\sum_{\ell\in\mathcal{S}\setminus j}|\bD_{j}^{\top}\bC_{\ell}|-\max_{j\in\mathcal{S}}|\bD_{j}^{\top}([-\zu+\vartheta]_r)|\nonumber\\
\label{eq:m2}\ge 1- (M-&1)\beta-\tau. 
\end{align}
\feop
Now we are ready to proof the Lemma-\ref{lem0}. The proof follows by the principle of mathematical 
induction. Suppose that the event $\Sigma$ occurs and $\bD$ satisfies \eqref{eq:d}. As 
\eqref{eq:d} holds, we always  find a threshold $\kappa$ satisfying \eqref{eq:k}. In the first iteration of CMUD $\zz^{(0)} = \zy$. The Lemma-\ref{lem2} implies that by setting a threshold $\kappa$ as in \eqref{eq:k} we can select at least one active code. Now 
suppose at $(\ell>1)$-th iteration $\mathbb{T}$ consists of the unique indices of active codes only and $\#\mathbb{T}=m$ with $m<M$. We have
\begin{align}
\zz^{(\ell-1)}=\sum_{j\in\mathcal{S}\setminus\mathbb{T}}\bC_j-\zu+\vartheta
\end{align}
By using the similar procedure in \eqref{eq:m1} and \eqref{eq:m2} it can be verified that
\begin{align}
\max_{j\not\in\{\mathcal{S}\setminus\mathbb{T}\}}\{|[\bD_{j}^{\top}\zz^{(\ell-1)}]_r|\}&\le (M-m)\beta+\tau\nonumber\\
\max_{j\in\{\mathcal{S}\setminus\mathbb{T}\}}\{|[\bD_{j}^{\top}\zz^{(\ell-1)}]_r|\}&\ge 1- (M-m-1)\beta-\tau.\nonumber
\end{align}
Clearly the threshold $\kappa$ satisfies $(M-m)\beta+\tau< \kappa < 1- (M-m-1)\beta-\tau$. Hence, CMUD will detect active code indices from $\{\mathcal{S}\setminus\mathbb{T}\}$. Now consider the iterate $p$, when all active codes are detected i.e., $\mathbb{T}=\mathcal{S}$. Then $\zz^{(p-1)}=-\zu+\vartheta$. Hence, as \eqref{eq:event} occurs, $|[\bD_{j}^{\top}\zz^{(p-1)}]_r|< \tau<\kappa$ and CMUD will not detect any further code.
\bibliographystyle{IEEEtran}
\bibliography{Letter,Tcsb}

\end{document}